\documentclass[11pt]{article}

\usepackage{amsmath}
\usepackage{amssymb}
\usepackage{graphicx}
\usepackage{xcolor}
\numberwithin{equation}{section}
\usepackage{array}
\usepackage{mathtools}
\usepackage{dsfont}
\usepackage{mathrsfs}
\usepackage{tikz}\usetikzlibrary{matrix,fit}
\usepackage{varwidth}
\usepackage{enumerate}
\usepackage[margin=1in]{geometry}
\usepackage{mathabx}
\usepackage{empheq}
\usepackage{hyperref}
\usepackage{setspace}
\usepackage[square,sort,comma,numbers]{natbib}

\newdimen\mytextwidth
\newcommand\rem[2][cyan!40!green]{\noindent\nobreak\hfil\penalty1000\hfilneg
\mytextwidth=\linewidth\advance\mytextwidth by 2mm%
\begin{tikzpicture}[baseline=-\the\dimexpr\fontdimen22\textfont2\relax]\node[outer sep=0pt,draw=black,fill=#1,fill opacity=1,text opacity=1,rectangle,rounded corners]{\begin{varwidth}{\mytextwidth}\textcolor{white}{#2}\end{varwidth}};
\end{tikzpicture}\allowbreak%
}

\newcommand{\dd}{\partial}
\newcommand{\bd}{\overline{\partial}}
\newcommand{\CP}{\mathds{CP}}
\newcommand{\CC}{\mathds{C}}

\renewcommand{\bar}{\overline}
\renewcommand{\tilde}{\widetilde}
\newcommand{\bnabla}{\bar{\nabla}}

\newcommand{\bea}{\begin{equation}}
\newcommand{\eea}{\end{equation}}
\newcommand{\bear}{\begin{eqnarray}}
\newcommand{\eear}{\end{eqnarray}}
\newcommand{\bearr}{\begin{eqnarray*}}
\newcommand{\eearr}{\end{eqnarray*}}

\newcommand{\appendixnumberline}[1]{Appendix.\space}

\let\oldappendix\appendix
\makeatletter
\renewcommand{\appendix}{%
  \addtocontents{toc}{\let\protect\numberline\protect\appendixnumberline}%
  \renewcommand{\@seccntformat}[1]{\large Appendix. }%
  \oldappendix
}
\makeatother

\usepackage{mdframed}

\newmdenv[
  topline=false,
  bottomline=false,
  rightline=false,
  linewidth=2pt,
  skipabove=\topsep,
  skipbelow=\topsep
]{siderules}

\newmdenv[
  topline=false,
  bottomline=false,
  linewidth=2pt,
  skipabove=\topsep,
  skipbelow=\topsep
]{siderulesright}

\newcommand\scalemath[2]{\scalebox{#1}{\mbox{\ensuremath{\displaystyle #2}}}}

\usepackage{setspace}
\usepackage{titlesec}

\titleformat*{\section}{\Large\bfseries}
\titleformat*{\subsection}{\large\bfseries}
\titleformat*{\subsubsection}{\large\bfseries}
\titleformat*{\paragraph}{\large\bfseries}
\titleformat*{\subparagraph}{\large\bfseries}

\begin{document}

\title{\vspace{-1.0cm} Deformed $\sigma$-models, Ricci flow and Toda field theories}
\author{Dmitri Bykov$^{1,2, 3}$\footnote{Emails:
bykov@mpp.mpg.de, bykov@mi-ras.ru} \;\;and Dieter L\"ust$^{1, 2}$\footnote{Email: dieter.luest@lmu.de}
\\ \\ 
{\small $^1$ Max-Planck-Institut f\"ur Physik, F\"ohringer Ring 6, D-80805 Munich, Germany}\\
{\small $^2$ Arnold Sommerfeld Center for Theoretical Physics,}\\ {\small Theresienstrasse 37, D-80333 Munich, Germany}\\
{\small $^3$ Steklov
Mathematical Institute of Russ. Acad. Sci.,}\\ {\small Gubkina str. 8, 119991 Moscow, Russia \;}}
\date{}

\begin{flushright}    
  {\small
    MPP-2020-59\\
    LMU-ASC 16/20
  }
\end{flushright}

{\let\newpage\relax\maketitle}

\maketitle

\vspace{-0.3cm}
\begin{siderulesright}
It is shown that the Pohlmeyer map of a $\sigma$-model with a toric two-dimensional target space naturally leads to the `sausage' metric. We then elaborate the trigonometric deformation of the $\CP^{n-1}$-model, proving that its $T$-dual metric is K\"ahler and solves the Ricci flow equation. Finally, we discuss a relation between flag manifold $\sigma$-models and Toda field theories.
\end{siderulesright}

\tableofcontents

\section{Introduction and main results}

The present paper is dedicated to two-dimensional $\sigma$-models with complex homogeneous target spaces and their deformations. By the results of~\cite{BykovNon, BykovSols, BykovZeroCurv, CYa}, these models are classically integrable, at least in the sense that their e.o.m. admit a zero-curvature representation. This is a generalization and extension of the results about $\sigma$-models with symmetric target spaces, in particular projective spaces $\CP^{n-1}$ and Grassmannians $Gr(k, n)$, that were widely studied in the 80's~\cite{Adda1, Adda2, Eichenherr0, Eichenherr1, Din1, Din2, Perelomov0, Morozov1, Perelomov1} (this is only an indicative list of references, as the literature on the subject is rather vast). The complex homogeneous spaces are torus bundles over flag manifolds~\cite{Wang}, and flag manifolds provide the basic and representative examples. The results of~\cite{CYa} (building on earlier work~\cite{CYa1, CYa2}) provide a way of constructing the so-called trigonometric and elliptic deformations of these models. The trigonometric deformation, also known as the $\eta$-deformation, was first considered by Cherednik~\cite{Cherednik}, extended and elaborated by Fateev~\cite{Fateev1} at the example of the principal chiral model, and substantially elucidated and generalized in the work of Klim\v{c}\'{\i}k~\cite{Klimcik1, Klimcik2, Klimcik3}. 
In their work~\cite{DMVq} Delduc, Magro and Vicedo constructed the $\eta$-deformation for the general case of symmetric target spaces, which is the setup relevant for the present paper. 
Further generalizations were proposed and elaborated later on 
in~\cite{Lukyanov,Matsumoto:2015ypa,Sfetsos:2015nya,Orlando:2016qqu,Osten:2016dvf,Georgiou:2016urf,DelducHoare, Sfetsos:2017sep,Lust:2018jsx,Orlando:2018kms,Georgiou:2018gpe,Klimcik4, Klimcik5}.
We should also mention that a lot of work has been dedicated to the study of the $AdS/CFT$ applications of such deformations, cf.~\cite{DMVAdS, Borsato, AFT, Tongeren,Itsios:2019izt}.

\vspace{0.3cm}\noindent
Below we mostly study the case of the trigonometrically-deformed $\CP^{n-1}$ model. In the case of $n=2$, and for a certain range of values of the deformation parameter $\eta$, the geometry is that of the so-called `sausage'~\cite{Onofri}. In section~\ref{Pohlmap} we provide an alternative derivation of the model. We start with a most general two-dimensional target space with a $U(1)$ isometry, in which case the geometry is described by a single function, which we call $g(\mu)$. It turns out that this function may be determined from the requirement of the integrability of the model. The first guess that comes to mind is that one could perform a dimensional reduction and require the integrability of the geodesic flow. For the two-dimensional geometry with a $U(1)$ symmetry this does not, however, impose any restrictions. Instead, we perform the Pohlmeyer reduction of the model~\cite{Pohlmeyer} (perhaps more properly called the Pohlmeyer map in the case at hand) and require the integrability of its mechanical version. The reduction itself rules out certain classical solutions, such as geodesics, which are `trivially' integrable anyway. The result of the analysis is that the integrable geometry turns out to be the `sausage', and a certain non-compact cousin thereof.

\vspace{0.3cm}\noindent
We view our derivation for $n=2$ as an empirical way of deducing the integrable geometry. For higher values of $n$ one could presumably follow the same route, but we leave this generalization for the future, and for the moment we utilize the machinery of~\cite{DMVq} to  construct the $\eta$-deformed version of the $\CP^{n-1}$ model. We note that the analysis of this geometry has been undertaken in the recent papers~\cite{LitvinovCP, FateevCP, DemulderCP}. In particular, it was noticed in~\cite{LitvinovCP} that it is useful to perform $T$-duality on all of the angles (the projective space, as well as its deformation, are toric manifolds), in order to get rid of the $B$-field. We will follow the same strategy here and mostly analyze the deformed model in the $T$-dual `frame'. Sometimes we refer to the dual geometry as $(\CP^{n-1})^{\vee}$. We derive, in explicit form, the metric of the dual model, which is the subject of proposition 2. The second result, formulated as proposition 3, is that the dual manifold is K\"ahler (which is also a statement that $\mathcal{N}=(2, 2)$ supersymmetry~\cite{DemulderCP} is preserved after $T$-duality). We obtain an explicit and compact expression for the K\"ahler potential, the main ingredient in this formula being the dilogarithm $\mathrm{Li}_2(z)$. 

\vspace{0.3cm}\noindent
In our paper we relate the $\eta$-deformed $\sigma$-models to the Ricci flow.
There is vast mathematical and physical literature on the Ricci flow. A nice review of the history of the developments around the Ricci flow equation, in particular in low dimensions, may be found in~\cite{Perelman}. Some other classic works on the subject (in $3$ and $2$ dimensions respectively) are~\cite{Hamilton1, Hamilton2}. One may also consult~\cite{Bakas2} as a concise general review of the theory of Ricci flow. Integrable structures of the Ricci flow equation for two-dimensional target spaces, related to the so-called `continual Toda system', were found in~\cite{Bakas1, Bakas2}. Similar integrable structures have been used to find solutions of the supergravity equations of motion~\cite{Cardoso}. Ricci flow has also been recently discussed in the context of the swampland distance conjectures~\cite{Kehagias}. 
Here we will show that, 
after a linear change of variables, the  dual metric of $(\CP^{n-1})^{\vee}$
satisfies the pure\footnote{I.e. without dilaton or an additional vector field} K\"ahler Ricci flow equation (proposition 4). The flow parameter, i.e. the time $\tau$ in this equation, is related to the deformation parameter $\eta$ in a simple way. The results of~\cite{Haagensen} then imply that the Ricci flow equation is satisfied in the original frame as well, this time with a dilaton and a non-zero $B$-field. The important role of the emerging dilaton is in the analysis of the effects of the Ricci flow on the metric: one has to take into account that the coordinates that were left intact by the Ricci flow in the $T$-dual frame start `running' (depending on $\tau$)  in the original frame, due to the presence of the dilaton. One has to take this into account in analyzing the asymptotic behavior of the metric under the Ricci flow.

\vspace{0.3cm}\noindent
In section~\ref{betagammadef} we point out that the recent work of Costello and Yamazaki~\cite{CYa} paves a way to constructing another integrable deformation of the $\CP^{n-1}$ model, and indeed of other models with complex homogeneous target spaces. This approach uses the language of $\beta\gamma$-systems. In fact, in this framework one quickly arrives at the flag manifold $\sigma$-models and related models introduced in~\cite{BykovNon, BykovSols, BykovZeroCurv}. As it has been shown in the recent work~\cite{DelducZT}, the Lax connections of a class of models closely related to these ones have ultra-local Poisson brackets. On the other hand, at least for $\CP^{n-1}$ the deformation based on the $\beta\gamma$-system formulation is different from the $\eta$-deformation, unless $n=2$. Perhaps this is due to the fact that the $\beta\gamma$-systems feature an anomaly in these cases,  and the corresponding models might not be well-defined quantum-mechanically. In view of this it is interesting, whether the ultralocal Poisson structure may be preserved for the $\eta$-deformed models when $n>2$ (for $n=2$ the answer is affirmative~\cite{Bytsko, Kotousov}).

\vspace{0.3cm}\noindent
Finally, in section~\ref{Todasec} we turn to the discussion of the flag manifold $\sigma$-models of~\cite{BykovNon, BykovSols, BykovZeroCurv}. We consider an analogue of `Pohlmeyer map' for these models. Whereas in the $\CP^1$-case this leads to a sin-Gordon model, in the case of complete flags $U(n)\over U(1)^n$ we obtain a field-theoretic incarnation of a  (periodic) Toda chain, interacting with some additional fields, which we call $U_{ij}$. When $U_{ij}=0$, we obtain a pure Toda field theory, and the map is the one that has been discussed in the math literature in the context of special harmonic maps (called primitive maps) to flag manifold target spaces~\cite{Bolton, Guest}. Our approach provides an embedding of these rather special maps into a full-fledged $\sigma$-model, with a non-topological $B$-field. If one drops the $U_{ij}$ fields, the zero-curvature representation for the $\sigma$-model reduces to the Lax pair of Flaschka-Manakov type.

\vspace{0.3cm}\noindent
\emph{Notation.} Throughout the paper the worldsheet $\Sigma$ will be assumed a Riemann surface, with complex coordinates $z, \bar{z}$. Derivatives with respect to these coordinates will be denoted $\dd:=\dd_z$ and $\bd:=\dd_{\bar{z}}$. We will also make use of worldsheet covariant derivatives, acting on sections of $T\mathcal{M}\big|_{\Sigma}$, i.e. on target space vectors restricted to the worldsheet, and these will be denoted $\nabla:=\nabla_z$ and $\bar{\nabla}:=\nabla_{\bar{z}}$. Similar notation is adopted for all other covariant derivatives, whose meaning is explained later on, i.e. $D, \bar{D}, \mathscr{D}, \bar{\mathscr{D}}$.

\section[Target space with a $U(1)$-isometry: the Pohlmeyer map]{Target space with a $U(1)$-isometry: the Pohlmeyer map}\label{Pohlmap}

To start with, we consider $\sigma$-models with a two-dimensional target space, i.e. maps $X: \Sigma\to \mathcal{M}_2$, where $\Sigma=\Sigma_2$ is the worldsheet  and $\mathcal{M}_2$ is the target-space with a $U(1)$-isometry. In particular, the metric on $\mathcal{M}_2$ has the form (in toric coordinates)
\begin{empheq}[box=\fbox]{align}
\hspace{1em}\vspace{1em}
\label{2dmetric}
ds^2=\sum\limits_{i, j=1}^2\, G_{ij}\,dX^i dX^j={1\over 4 \,g(\mu)}\,d\mu^2+g(\mu)\, d\phi^2\,.
\hspace{1em}
\end{empheq}
Here $g^{-1}=H_{\mu\mu}$ may be understood as the second derivative of a `symplectic potential' $H(\mu)$, $\mu$ being the moment map for the $U(1)$-action. In this section we will not need this detail, but we will return to it in Sec.~\ref{Tdualtoric} in the discussion of higher-dimensional toric manifolds.

\vspace{0.3cm}\noindent
We wish to decide, for which geometries (i.e. for which functions $g(\mu)$) the $\sigma$-model associated with the above metric is classically integrable. The first thing to note is that the direct mechanical reduction of the model does not give any clues, as the resulting mechanical system is integrable for any $g(\mu)$. Indeed, it possesses two integrals of motion: the energy and the angular momentum (corresponding to the shifts of $\phi$). For this reason we will take another route and look at the generalized Pohlmeyer reduction of the model.

\vspace{0.3cm}\noindent
Following the original approach of Pohlmeyer~\cite{Pohlmeyer}, we may set
\bea\label{pohlunitcond}
G_{ij}\,\dd X^i \dd X^j=1=G_{ij}\,\bd X^i \bd X^j\,,
\eea
using conformal invariance of the $\sigma$-model action. An analogue of the Cauchy-Schwarz inequality for the complex vector $\dd X^i$ implies $G_{ij}\,\dd X^i \bd X^j\geq 1$, therefore
\bea\label{psidef}
G_{ij}\,\dd X^i \bd X^j=\cosh{\chi}
\eea
The equations of motion of the $\sigma$-model have the form $ \bar{\nabla}\nabla X^j=0$.

\vspace{0.3cm}\noindent
We will now derive the equation of motion for the field $\chi$. In the undeformed case, when $\mathcal{M}_2=S^2$ is a sphere, this equation is the sinh-Gordon equation, as observed by Pohlmeyer\footnote{More precisely, Pohlmeyer considered the model on a worldsheet with Minkowski signature, in which case one gets the sin-Gordon model. This is due to the fact that the vectors $\dd_\pm X$ are real, whereas $\dd X$ and $\bd X$ are complex, which leads to obvious differences in the two derivations.}. 

\vspace{0cm}\noindent
\begin{siderules}
{\large\textbf{Proposition 1.}} Suppose the e.o.m. of the $\sigma$-model with the metric~(\ref{2dmetric}), as well as the Pohlmeyer conditions~(\ref{pohlunitcond}), are satisfied. Then $\chi(z, \bar{z}), \;\mu(z, \bar{z})$ satisfy the~equations
\bear
&&\bd\dd\chi-2g''(\mu)\,\sinh{\chi}=0\\
&&\dd\bd \mu-2\,g'(\mu)\, \cosh{\chi}=0\,.
\eear
\end{siderules}

\vspace{0.3cm}\noindent
{\large \textbf{Proof.}} 
First of all, let us assume that the vectors $\dd X^i$ and $\bd X^i$ are linearly independent at generic points. In this case, since the target space is two-dimensional, we may decompose any vector field $v\big|_\Sigma$ in this basis. For example,
\bea\label{nablaz2}
\nabla^2 X^i=a_z\,\nabla X^i+b_z\,\bnabla X^i\,.
\eea
To calculate the coefficients, we take scalar products with the respective vectors and use the conditions~(\ref{pohlunitcond}), as well as the e.o.m. for $X^i$ (the brackets $\langle \bullet, \bullet\rangle$ indicate the scalar product w.r.t. the metric $G_{ij}$):
\bear
&&\langle\nabla X, \nabla^2 X \rangle={1\over 2}\dd\langle\nabla X, \nabla X \rangle=0=a_z+b_z\,\cosh{\chi}\\
&&\langle\bnabla X, \nabla^2 X \rangle=\dd(\cosh{\chi})=a_z\,\cosh{\chi}+b_z\,.
\eear
One finds
\bea
a_z=\dd \chi \cdot\frac{\cosh{\chi}}{\sinh{\chi}},\quad\quad b_z=-\frac{\dd \chi}{\sinh{\chi}}\,.
\eea
Analogously
\bear\label{nablazbar2}
&&\bnabla^2 X^i=a_{\bar{z}}\,\nabla X^i+b_{\bar{z}}\,\bnabla X^i\,,\\
&& \textrm{where}\quad\quad a_{\bar{z}}=-\frac{\bd \chi}{\sinh{\chi}},\quad\quad b_{\bar{z}}=\bd \chi \cdot\frac{\cosh{\chi}}{\sinh{\chi}}\,.
\eear
To derive the e.o.m. of $\chi$, we compute the mixed derivative $\bd \dd$ of the definition $\langle\bnabla X, \nabla X \rangle=\cosh{\chi}$:
\bea
\cosh{\chi}\,\bd\chi\, \dd\chi+\sinh{\chi}\,\dd\bd\chi=\langle\bnabla^2X, \nabla^2 X\rangle+\langle\bnabla X, \bnabla\nabla^2 X\rangle
\eea
From (\ref{nablaz2}) and (\ref{nablazbar2}) we see that the first terms in both sides are equal. On the other hand, for the last term we use the following identity:
\bea
\bnabla\nabla^2 X^i=[\bnabla, \nabla] \nabla X^i+\nabla \bnabla\nabla X^i=R^i_{\;mkn}\;\bnabla X^k\,\nabla X^m\,\nabla X^n\,,
\eea
where $R$ is the Riemann tensor of the metric~(\ref{2dmetric}). For a two-dimensional  target space with  complex coordinate $w$ and line element $ds^2=\Omega \,dw\,d\bar{w}$ we get $R_{w\bar{w}w\bar{w}}={\Omega\over 2} \,\bar{\dd}\dd\log{\Omega}$ and
\bea
\langle\bnabla X, \bnabla \nabla^2 X\rangle=-{1\over 8}\Omega^2 \left(\bd w \dd \bar{w}-\dd w \bd \bar{w}\right)^2\,R\,
\eea
where $R$ is the scalar curvature $R=-4\,\Omega^{-1}\,\bar{\dd}\dd\log{\Omega}$. Since ${\Omega\over 2} \left(\bd w \dd \bar{w}+\dd w \bd\bar{w}\right)=\cosh{\chi}$, taking into account~(\ref{pohlunitcond}), we find
\bea
\langle\bnabla X, \bnabla\nabla^2 X\rangle=-{1\over 2}\,\sinh^2{\chi}\,R\,,
\eea
therefore the e.o.m. for $\chi$ takes the form
\bea\label{psieq}
\bd\dd\chi+{1\over 2}\,R\,\sinh{\chi}=0\,.
\eea
In fact so far we have not made use of the toric structure~(\ref{2dmetric}) of the target space. Now, for the toric metric~(\ref{2dmetric}) the scalar curvature is
\bea
R=-4\,\frac{d^2 g}{d\mu^2}\,.
\eea
For example, in the case of a sphere $g={1\over 2}(1-\mu^2)$, so that the scalar curvature $R=4$ (compatible with the Gauss-Bonnet formula ${1\over 4\pi}\,\int\,{1\over 2}\,d\mu\,d\phi\,R=2$), and one arrives at the sinh-Gordon equation. In all other cases, however, $R=R(\mu)$ is a non-constant function, and therefore the equation for $\chi$ involves the target-space field $\mu$. One therefore needs to write out the e.o.m. for $\mu$, which easily follows from the $\sigma$-model action with the metric~(\ref{2dmetric}):
\bea\label{mueq}
\dd\bd \mu-2\,g'\, \cosh{\chi}=0\,.
\eea
Here we also used the definition~(\ref{psidef}) of $\chi$.
(For completeness we also record the e.o.m. for the remaining target-space variable $\phi$:
$
\bd(g(\mu)\,\dd \phi)+\dd (g(\mu)\,\bd\phi)=0\,.
$) \dotfill $\blacksquare$

\vspace{0.3cm}\noindent
Now, the equations~(\ref{psieq}) and (\ref{mueq}) follow from a single Lagrangian
\bea\label{alagr}
\mathscr{L}={1\over 2 } \dd\mu \bd \mu+{1\over 2} \dd\chi \bd \chi+f(\mu)\,\cosh{\chi}
\eea
if the following conditions are satisfied: $f=-{R\over 2}=2g''$, $f'=2 g'$, which leads to the equation  $f''-f=0$. As a result,  $f=2a_1 \,e^\mu+2a_2\, e^{-\mu}$. The function $g(\mu)$ entering the metric~(\ref{2dmetric}) is
\bea
g(\mu)=b+a_1\, e^{ \mu}+a_2\, e^{-\mu}\,,
\eea
where $a_1, a_2, b$ are apriori arbitrary constants. Interestingly, for certain choices of these parameters the corresponding metric corresponds to the so-called `sausage' (an elongated sphere). For the moment let as assume that $\mathrm{sgn}(a_1)=\mathrm{sgn}(a_2)$, in which case, by a rescaling and shift of $\mu$, we can bring the function $g$ to the form
\bea
g(\mu)=b+a\,\cosh{\mu}\,.
\eea
If we wish $\mathcal{M}_2$ to be of finite volume, we should assume that the function $g(\mu)$ has two zeros, $g(\pm \mu_0)=0$. Provided the angle $\phi$ has periodicity $2\pi$, the non-singularity at these points requires that $g'(\pm \mu_0)=\mp 1$. This leads to the conditions $b+a\,\cosh{\mu_0}=0,\; a\,\sinh{\mu_0}=-1$, i.e.
\begin{empheq}[box=\fbox]{align}
\hspace{1em}\vspace{1em}
\label{gsausage}
g(\mu)=\frac{\cosh{\mu_0}-\cosh{\mu}}{\sinh{\mu_0}}\,,\quad\mu\in[-\mu_0, \mu_0]\,.
\hspace{1em}
\end{empheq}
The metric~(\ref{2dmetric}) with this function $g(\mu)$ is the so-called `sausage' geometry~\cite{Onofri}, which is a deformation of the round sphere $S^2$. Here $\mu_0$ is the deformation parameter, and the round sphere metric is recovered in the limit $\mu_0\to0, \mu=\mu_0 \tilde{\mu}$, with $\tilde{\mu}\in [-1, 1]$. 

\vspace{0.3cm}\noindent
We obtained the above results by requiring that the $\mu, \chi$-equations follow from a single Lagrangian. One could argue that this requirement is too stringent, since it is known that typically Pohlmeyer's reduction may lead to non-Lagrangian equations~\cite{Grigoriev}. A justification is that our goal is to understand, whether the original (and hence the resulting) model is integrable. One simple check of this consists in performing a mechanical reduction. For the mechanical system, however, the requirement that the two equations follow from a single Lagrangian is the same as the requirement that the system possesses an integral of motion, the energy. Miracuosly the second integral of motion, necessary for a system with two degrees of freedom, appears automatically. Indeed, let us write out the Lagrangian~(\ref{alagr}) for the function $f(\mu)=2a\,\cosh{\mu}$:
\bear\label{muchiLagr}
&&\mathscr{L}={1\over 2} \dd\mu \bd \mu+{1\over 2} \dd\chi \bd \chi+2a\,\cosh{\mu}\,\cosh{\chi}\,=\\
&&=\left({1\over 2} \dd\tilde{\mu} \,\bd \tilde{\mu}+a\,\cosh{(\sqrt{2}\tilde{\mu})}\right)+\left({1\over 2} \dd\tilde{\chi}\, \bd \tilde{\chi}+a\,\cosh{(\sqrt{2}\tilde{\chi})}\right)\,,\\
&&\textrm{where}\quad\quad \tilde{\mu}={\mu+\chi\over \sqrt{2}},\quad\quad\tilde{\chi}={\mu-\chi\over \sqrt{2}}\,.
\eear
We have arrived at a system of two independent fields. Their respective energies are conserved, leading to the second integral of motion in the mechanical setting.

\vspace{0.3cm}\noindent
It would be interesting to extend this analysis to higher-dimensional homogeneous spaces, whose Pohlmeyer reductions have been studied extensively, cf.~\cite{Eichenherr, Grigoriev, Miramontes} and references therein.

\vspace{0.3cm}\noindent
\emph{Comment.} In the derivation above we assumed $\mathrm{sgn}(a_1)=\mathrm{sgn}(a_2)$. If, on the contrary, $\mathrm{sgn}(a_1)=-\mathrm{sgn}(a_2)$, after a shift of $\mu$ the function $g$ is found to be $g(\mu)=\frac{\sinh{\mu}-\sinh{\mu_0}}{\cosh{\mu_0}}\,,\;{\mu\geq \mu_0}\,.$ This is clearly an unbounded geometry. In the limit $\mu_0\to0, \mu=\mu_0 \tilde{\mu}$, with $\tilde{\mu}\in [1, \infty)$, one obtains the flat space $\mathbb{R}^2$.

\subsection{The round sphere: $\mathcal{M}_2=S^2$}
In the case of a sphere, when $g=g_0:={1\over 2}(1-\mu^2)$, the equations for $\chi, \mu, \phi$ take the form:
\bear\label{chieqlin}
&&\dd\bd\chi+2\sinh{\chi}=0\,.\\
&&\dd\bd\mu+2\mu\,\cosh{\chi}=0\,.\\
&&\bd(g_0(\mu)\,\dd \phi)+\dd (g_0(\mu)\,\bd\phi)=0\,.
\eear
Therefore the only essentially nonlinear equation is the first one, and the remaining two are merely a set of nested linear equations.

\vspace{0.3cm}\noindent
Let us now derive the Lax pair of the sinh-Gordon model from the Lax pair of the $\CP^1$-model. The latter arises from the condition of zero curvature of the family of connections 
\bear\label{pohlflat1}
&&A_u={1-u\over 2} K_z dz+{1-u^{-1}\over 2} K_{\bar{z}} d\bar{z}\,,\quad\textrm{where}\\
&& K_z=g\begin{pmatrix}
0 & 2\,(J_{12})_z \\
0 & 0 
\end{pmatrix}g^\dagger,\quad\quad K_{\bar{z}}=-K_z^\dagger=g\begin{pmatrix}
0 & 0 \\
2\,(J_{21})_{\bar{z}} & 0 
\end{pmatrix}g^\dagger
\eear
Here $J_{12}=\bar{u}_1\circ du_2$ and $J_{21}=\bar{u}_2\circ du_1$ are components of the Maurer-Cartan current $J=g^\dagger dg$, where the matrix $g=(u_1, u_2)$, and we assume $\mathrm{det}(g)=1$. Upon a gauge transformation by the group element $g$ we obtain
\bear
\mathscr{A}_u=
\begin{pmatrix}
(J_{11})_z dz+(J_{11})_{\bar{z}} d\bar{z} & u\, (J_{12})_z dz+(J_{12})_{\bar{z}} d\bar{z} \\
u^{-1}\, (J_{21})_{\bar{z}} d\bar{z}+(J_{21})_{z} dz & -(J_{11})_z dz-(J_{11})_{\bar{z}} d\bar{z}
\end{pmatrix}
\eear
Pohlmeyer's condition~(\ref{pohlunitcond}) may be formulated as $(J_{12})_z (J_{21})_z={1\over 2}$. We set $(J_{12})_z:={1\over \sqrt{2}}e^y$. Due to our parametrization of $\CP^1\simeq {SU(2)\over U(1)}$, we have the gauge group $U(1)$ acting on $J_{12}$ by phase rotations $J_{12}\to e^{i \alpha}\,J_{12}$, therefore we can choose a gauge, in which $y$ is real. Then $(J_{21})_z={1\over \sqrt{2}}e^{-y}, (J_{12})_{\bar{z}}=-((J_{21})_z)^\ast=-{1\over \sqrt{2}}e^{-y}, (J_{21})_{\bar{z}}=-((J_{12})_z)^\ast=-{1\over \sqrt{2}}e^{y}$. From the e.o.m. $D_{\bar{z}}(J_{12})_z=(\dd_{\bar{z}}+2(J_{11})_{\bar{z}})(J_{12})_z =0$ we find $(J_{11})_{\bar{z}}=-{1\over 2} \dd_{\bar{z}}y $. The connection takes the form
\bea
\mathscr{A}_u=\begin{pmatrix}
{1\over 2}\,\dd_z y\, dz- {1\over 2}\,\dd_{\bar{z}} y \,d\bar{z} & {1\over \sqrt{2}}\,u\, e^y dz-{1\over \sqrt{2}}\,e^{-y} d\bar{z} \\
-{1\over \sqrt{2}}\,u^{-1}\, e^{y} d\bar{z}+{1\over \sqrt{2}}\,e^{-y} dz &  {1\over 2}\,\dd_{\bar{z}} y \,d\bar{z}-{1\over 2}\,\dd_z y\, dz
\end{pmatrix}
\eea
The zero-curvature equation $dA_u+A_u\wedge A_u=0$ then results in the sinh-Gordon equation $\dd\bd y+\,\mathrm{sinh}(2y)=0$. Setting $y={\chi\over 2}$, we return to~(\ref{chieqlin}). In section~\ref{genTodasec} we will generalize this construction to the case of complete flag manifolds in arbitrary dimension, arriving at a Toda field theory as the relevant counterpart of the sinh-Gordon model.

\section{The deformed $\CP^{n-1}$-model}\label{etadefsec}

Before starting this section, we point out that the $\eta$-deformation of the $\CP^{n-1}$-model has been recently considered in the works~\cite{LitvinovCP, FateevCP, DemulderCP}. We will comment below on the relation of their results to our discussion.

\subsection{$\eta$-deformation}

The trigonometric deformation (often called $\eta$-deformation) of the principal chiral model was introduced by Klim\v{c}\'{\i}k~\cite{Klimcik1, Klimcik2, Klimcik3}, and the result was generalized by Delduc-Magro-Vicedo to the case of symmetric spaces in~\cite{DMVq}. To describe the latter prescription, first we introduce some notation. Let $g\in G$ be the group element, $J=g^{-1}dg$ the standard Maurer-Cartan current, $\mathfrak{g}=\mathfrak{h}\oplus \mathfrak{m}$ the decomposition of the Lie algebra of the group $G$, and $\Pi_{\mathfrak{h}}, \Pi_{\mathfrak{m}}$ the projectors on the corresponding subspaces. The $\eta$-deformed model of~\cite{DMVq} is formulated as follows:
\bear
&&\mathscr{L}=\mathrm{Tr}\left((J_{\bar{z}})_\mathfrak{m}\,\frac{1}{1-\eta\,\Pi_\mathfrak{m}\,R_g\,\Pi_\mathfrak{m}}\,(J_z)_\mathfrak{m}\right),\\
&& R_g=\mathrm{Ad}_{g^{-1}}\circ\mathcal{R}\circ\mathrm{Ad}_g\,,
\eear
where $\mathcal{R}\in \mathrm{End}(\mathfrak{g})$ is defined as follows: $\mathcal{R}=i\,(\pi_+-\pi_-)$, $\pi_\pm$ being the projections on the upper/lower-triangular matrices. First of all, we get rid of the denominator by introducing auxiliary fields\footnote{Since $J_{\bar{z}}=-J_z^\dagger$ and $M_{\bar{z}}=M_z^\dagger$, the sum of the first two terms in the Lagrangian is in fact imaginary. This is due to the fact that we are working in Euclidean signature.} $M_z, M_{\bar{z}}=M_z^\dagger \in \mathfrak{g}_\CC$:
\bea
\mathscr{L}=\mathrm{Tr}\left((J_{\bar{z}})_\mathfrak{m}\,M_z\right)+\mathrm{Tr}\left(M_{\bar{z}}(J_z)_\mathfrak{m}\right)+   \mathrm{Tr}\left(\mathrm{Ad}_{g}((M_{\bar{z}})_\mathfrak{m})\,\left(1-\eta\,\mathcal{R}\,\right)\,\mathrm{Ad}_g((M_z)_\mathfrak{m})\right)\,.
\eea
Now, we wish to formulate the model in terms of the homogeneous coordinates $U_i$, rather than in terms of the unitary group element $g$. The relation between the two is as follows. Introduce the `pseudo-metric' $\Lambda=\mathrm{Diag}(1, -1, \ldots , -1)$ and the group element
\bea\label{Ggroup}
G=g\Lambda g^\dagger\,,\quad\quad (G^\dagger=G, \quad G^2=\mathds{1})\,.
\eea
The virtue of $G$ is that it is gauge-invariant w.r.t. the gauge group $H=U(1)\times U(n-1)$ (the stabilizer of the quotient space $\CP^{n-1}=\frac{U(n)}{U(1)\times U(n-1)}$). Such gauge-invariant parametrizations were first considered in the context of $\sigma$-models in~\cite{Eichenherr1}. In fact, the map $g\to G$ is an explicit form of Cartan's embedding $\CP^{n-1}\hookrightarrow U(n)$, which can be written out in the homogeneous coordinates, as desired:
\bea
G=\mathds{1}-2\,\frac{U\otimes \bar{U}}{\|U\|^2}\,.
\eea
It follows from the definition~(\ref{Ggroup}) that
\bea
\mathcal{J}:=G^\dagger dG=-\,g\,\left(g^\dagger dg-\Lambda g^\dagger dg \Lambda \right)\,g^\dagger=-2\,g\,J_{\mathfrak{m}}\,g^\dagger\,.
\eea
Since $\Pi_{\mathfrak{m}}=\frac{1-\mathrm{Ad}_\Lambda}{2}$, we can write $(M_z)_\mathfrak{m}={1\over 2}\left(M_z-\Lambda M_z \Lambda\right)$. If we now make a field redefinition $M_z \to g^\dagger M_z g$, we can recast the Lagrangian in the form
\bear\nonumber
&&\mathscr{L}=-{1\over 2}\mathrm{Tr}\left((\mathcal{J}_{\bar{z}})_\mathfrak{m}\,\tilde{M}_z\right)-{1\over 2}\mathrm{Tr}\left(\tilde{M}_{\bar{z}}(\mathcal{J}_z)_\mathfrak{m}\right)+   \mathrm{Tr}\left((\tilde{M}_{\bar{z}})\,\left(1-\eta\,\mathcal{R}\,\right)\,(\tilde{M}_z)\right)\,,\\
&&\textrm{where}\quad\quad \tilde{M}_z={1\over 2}\left(M_z-G M_z G\right)\,.
\eear
The last line contains a projector, which projects on matrices $M$ satisfying $M+G M G=0$. Such matrices can be described as follows:

\noindent
\begin{siderules}
{\large \textbf{Lemma 1.}} A matrix $M$ satisfying $M+G M G=0$ can be parametrized as follows:
\bea\label{MVW}
M=U\otimes \bar{W}+V\otimes \bar{U}\,
\eea
where $\bar{W}\circ U=\bar{U}\circ V=0$, i.e. $V, W$ are two vectors orthogonal to $U$.
\end{siderules}

\vspace{0.3cm}\noindent
{\large \textbf{Proof.}} We write out the equation explicitly:
\bea
M+G M G=2M+4 \,\frac{U\otimes \bar{U}\,(\bar{U} M U)}{\|U\|^4}-2\,\frac{U\otimes \bar{U}M}{\|U\|^2}-2\,\frac{MU\otimes \bar{U}}{\|U\|^2}=0\,.
\eea
Multiplying by $\bar{U}$, one finds $\bar{U} M U=0$, therefore $M=\frac{U\otimes \bar{U}M}{\|U\|^2}+\frac{MU\otimes \bar{U}}{\|U\|^2}$. Denoting $V=\frac{MU}{\|U\|^2}$ and $\bar{W}=\frac{\bar{U}M}{\|U\|^2}$, we bring $M$ to the desired form. The vectors $V$ and $W$ are orthogonal to $U$ due to $\bar{U} M U=0$. Conversely, a matrix of the form~(\ref{MVW}) satisfies $M+G M G=0$. \dotfill $\blacksquare$

\vspace{0.7cm}\noindent
Next we need to substitute $M$ from~(\ref{MVW}) in place of $\tilde{M}_z$ and $M^\dagger$ in place of $\tilde{M}_{\bar{z}}$ accordingly. To this end, it is useful to write an explicit expression for the current $\mathcal{J}$:
\bea
{1\over 2}\mathcal{J}={1\over 2}G^\dagger dG=\frac{U\otimes D\bar{U}-DU\otimes \bar{U}}{\|U\|^2}\,,\quad\quad DU=dU-\frac{\bar{U}\circ dU}{\|U\|^2}\,U\,.
\eea
 Besides, we introduce the constraints $\bar{W}\circ U=0$ and $\bar{U}\circ V=0$ using the complex Lagrange multipliers $\mathcal{B}$ and $\mathcal{C}$:
\bear\label{Lagrdefrho}
&&\mathscr{L}=\left(\bar{U}(\mathcal{D}_{\bar{z}}V)\right)+\left(\bar{U} (\mathcal{D}_{z}W)\right) -\left((\mathcal{D}_{\bar{z}}\bar{W})U\right)-\left((\mathcal{D}_{z}\bar{V})U\right)+\\ \nonumber&&+  \mathrm{Tr}\left(W\otimes \bar{U} \,\rho(U\otimes \bar{W})\right)+\mathrm{Tr}\left(U\otimes \bar{V} \,\rho(V\otimes \bar{U})\right)+\\ \nonumber
&& +  \mathrm{Tr}\left(U\otimes \bar{V} \,\rho(U\otimes \bar{W})\right)+\mathrm{Tr}\left(W\otimes \bar{U} \,\rho(V\otimes \bar{U})\right)\,,\\ \nonumber
&&\textrm{where}\quad\quad \mathcal{D}_{z}W=\dd_z W-i \,\mathcal{B}\,W,\quad\quad \mathcal{D}_{\bar{z}}V=\dd_{\bar{z}} V-i\, \mathcal{C}\,V\,,
\eear
where we have introduced the notation $\rho=1-\eta\,\mathcal{R}$. As elaborated in~\cite{DemulderCP}, this is really an example of generalized K\"ahler geometry. In particular, the quadratic form in the last two lines of~(\ref{Lagrdefrho}) encodes the generalized K\"ahler metric~\cite{Gualtieri}. 

\vspace{0.3cm}\noindent
The matrix $\rho$ is closely related to the (trigonometric) classical $r$-matrix. We recall a few facts about the latter. In the analysis of the Yang-Baxter equation (see App.~\ref{CYBEapp}) one usually assumes that the $r$-matrix takes values in $\mathfrak{g}\otimes \mathfrak{g}$, however for the purposes of this paper it is more useful to think of it as taking values in $\mathrm{End}(\mathfrak{g})$, the two definitions being related by raising/lowering an index using the Killing metric on $\mathfrak{g}$. The $r$-matrix then has the form (we assume $\mathfrak{g}\simeq \mathfrak{su}_n$)
\bear\label{trigrmatrix}
&&r_s=\alpha_s\,\pi_++\beta_s \,\pi_-+\gamma_s\,\pi_0\,,\\
&&\alpha_s=\frac{1}{1-s},\quad\quad \beta_s=\frac{s}{1-s},\quad\quad \gamma_s={1\over 2}\,\frac{1+s}{1-s}\,,
\eear
where $\pi_\pm$ are projections on the upper/lower-triangular matrices, and $\pi_0$ is the projection on the diagonal. The rational limit is achieved by setting $s=e^{-v}$ and taking the limit $v\to 0$, in which case $r_s\to \frac{\mathds{1}}{v}$. Now, the relation of $r_s$ to $\rho$ is $r_s={\rho\over 2i\eta}$, if one assumes the following relation between the parameters:
\bea\label{qeta}
s=\frac{1-i\,\eta}{1+i\,\eta}\,.
\eea
In fact, this suggests introducing the notation $\eta=\tan{({t_n\over 2})}$, in which case $s=e^{-i \,t_n}$. We will be using this parametrization later on.

\vspace{0.3cm}\noindent
We return to the Lagrangian~(\ref{Lagrdefrho}). There are several ways to proceed from this starting point. One can integrate out the fields $V, W$ (the quadratic part is a Hermitian form in the variables $(V, \bar{W}), (\bar{V}, W)$) and get a deformed $\CP^{n-1}$ Lagrangian that depends on the $U$-fields. The calculation along these lines seems rather formidable and leads to a generalized K\"ahler manifold~\cite{DemulderCP}. As it has already been observed in~\cite{LitvinovCP}, a simpler answer can be obtained if one performs $T$-duality in all the angular directions of the deformed space. The reason is that the $B$-field of the above model has the form $B=\sum\,b_i\wedge d\phi_i$, where $b_i$ are some one-forms that do not depend on $\phi_j$. In this case, according to the Buscher rules~\cite{Buscher} (for a review of $T$-duality see also~\cite{Plauschinn}), $T$-duality along all the directions $\phi_i$ eliminates the $B$-field altogether. We will also follow this path, but first we recall how $T$-duality works for K\"ahler manifolds.

\subsection{$T$-duality for toric K\"ahler manifolds}\label{Tdualtoric}

Although the $\eta$-deformed $\CP^{n-1}$ manifold is not K\"ahler in general (unless $n=2$), let us first recall that, in the K\"ahler setup, $T$-duality may be implemented by performing a Legendre transform on the K\"ahler potential. This is especially easy to see in the case of a toric variety (see~\cite{Guillemin, Abreu} for background on toric K\"ahler manifolds), and when $T$-duality is performed on all angles simultaneously. Indeed, suppose we have a K\"ahler manifold of complex dimension $n$, with complex coordinates $W_1, \ldots, W_n$. Assume that the torus acts by shifting $W_k \to W_k +i\, \delta \phi_k$, i.e. the imaginary parts of $W_k$ are the angles. On a toric manifold, the K\"ahler potential may be then assumed to be of the form
\bea\label{Kahtoric}
\mathcal{K}=\mathcal{K}(\underbracket[0.6pt][0.6ex]{W_1+\bar{W}_1}_{:=t_1}, \ldots, \underbracket[0.6pt][0.6ex]{W_n+\bar{W}_n}_{:=t_n})\,.
\eea
The line element is
\bea\label{metrtoric1}
ds^2=\sum\limits_{j,k=1}^n\,\left({1\over 4}\frac{\dd^2 \mathcal{K} }{\dd t_j \dd t_k}\,dt_j\,dt_k+\frac{\dd^2 \mathcal{K} }{\dd t_j \dd t_k}\,d\phi_j\,d\phi_k\right)\,.
\eea
It is also useful to define the moment map variables $\mu_k={\dd  \mathcal{K}\over \dd t_k}$, and the dual potential $H(\mu_1, \ldots, \mu_n)$, also known as the symplectic potential. It is related to $\mathcal{K}$ by a Legendre transform: $H(\mu_1, \ldots, \mu_n)=\sum\limits_{j=1}^n\,\mu_j\,t_j-\mathcal{K}(t_1, \ldots, t_n)$. An elementary property of the Legendre transform is that the Hessian of $H$ is inverse to the Hessian of $K$, i.e. suppressing the matrix indices we may write ${\dd^2 H\over \dd \mu^2}=({\dd^2 K\over \dd t^2})^{-1}$. In terms of these new variables the metric takes the form
\bea\label{metrtoric2}
ds^2=\sum\limits_{j,k=1}^n\,\left({1\over 4}\frac{\dd^2 H }{\dd \mu_j \dd \mu_k}\,d\mu_j\,d\mu_k+\left(\frac{\dd^2 H }{\dd \mu^2}\right)^{-1}_{jk}\,d\phi_j\,d\phi_k\right)\,.
\eea
We note in passing that, for $n=1$, this gives essentially the formula~(\ref{2dmetric}) for the metric of the two-dimensional space that we studied in Sec.~\ref{Pohlmap}. One of the virtues of the symplectic potential $H$ is that its domain of definition is of geometric significance -- it is the moment polytope of the manifold.

\vspace{0.3cm}\noindent
If we now perform $T$-duality on all angles $\phi_k$ simultaneously, this will not generate any $B$-field, but the matrix in the second term of~(\ref{metrtoric2}) will be inverted. In this case the metric takes the form~(\ref{metrtoric1}), but with the replacement $K\leftrightarrow H, t_k \leftrightarrow \mu_k$. In other words, $T$-duality works by interchanging the K\"ahler potential with its Legendre-dual, the symplectic potential.

\vspace{0.3cm}\noindent
From the point of view of supersymmetric theory, K\"ahler target spaces correspond to $\mathcal{N}=(2, 2)$ worldsheet supersymmetry. The complex coordinates correspond to chiral superfields, and partial $T$-dualities replace chiral superfields by twisted chiral superfields~\cite{RocekVerlinde, LindstromRocek}. In general, the presence of both chiral and twisted chiral superfields leads to a non-topological $B$-field. However, if we $T$-dualize all variables, all chiral superfields are replaced by twisted chiral superfields, which is effectively the same as having all chiral superfields and leads again to K\"ahler geometry, described by the dual potential.

\vspace{0.3cm}\noindent
This leads us then to the following curious question. We start with the $\eta$-deformed $\CP^{n-1}$-geometry, which is a toric manifold and has, in general, a non-topological $B$-field. Next we $T$-dualize all angles and get rid of the $B$-field completely. Since we expect $\mathcal{N}=(2, 2)$ supersymmetry to be preserved, the resulting manifold should be K\"ahler, again with $n$ commuting isometries. If its K\"ahler potential was of the form~(\ref{Kahtoric}), the metric would have the form~(\ref{metrtoric1})-(\ref{metrtoric2}), but in this case the inverse $T$-duality would not generate any $B$-field and therefore would not lead us to the original geometry. So how is this apparent puzzle resolved? In section~\ref{TdualKahler} we will analyze the $T$-dual K\"ahler geometry in detail, but the ultimate answer will be that, due to the fact that this manifold is no longer simply-connected, the moment map variables $\mu_k$ are not well-defined (only the one-forms $d\mu_k$ are). As a result, the $T$-dual metric cannot really be brought to the form~(\ref{metrtoric1})-(\ref{metrtoric2}). Before passing over to this derivation, we briefly recall the symplectic geometry of the $\CP^{n-1}$ space. This will be useful for the following sections, since essentially the same variables may be used for the description of the deformed geometry.

\subsubsection{The undeformed $\CP^{n-1}$ geometry and its $T$-dual}

The complex projective space is defined as the set of equivalence classes of $n$-tuples of complex numbers $(Q_1, \ldots, Q_n )\in \CC^n\setminus\{0\}$ w.r.t. to the $\CC^\times$-action $Q_i \to \lambda \,Q_i$\,. One may also describe it using a symplectic (K\"ahler) quotient of $\CC^n$ w.r.t. a $U(1)$-action that preserves the standard symplectic form. This leads to the K\"ahler potential that in homogeneous coordinates takes the form
\bea
\mathcal{K}
=\log{\left(\sum\limits_{j=1}^n |Q_j|^2\right)}\,.
\eea
In the previous section we assumed an additive, rather than multiplicative, action of the torus, therefore we set $Q_j=e^{W_j}$, so that $\mathcal{K}=\log{\left(\sum\limits_{j=1}^n e^{\tilde{t}_j}\right)}$. We have marked the $t$-variables with a tilde, since ultimately we will be interested in the $T$-dual K\"ahler potential, and we reserve the unmarked $t$-variables for that case. The moment map variables are then $\mu_j={\dd \mathcal{K}\over \dd \tilde{t}_j}=\frac{e^{\tilde{t}_j}}{\sum\limits_{j=1}^n e^{\tilde{t}_j}}$. The symplectic potential has the form
\bea\label{GCPN}
H=\sum\limits_{j=1}^n\,\mu_j\,\log{\mu_j},\quad\quad\textrm{where}\quad \mu_j\geq 0 \quad \textrm{and}\quad \sum\limits_{j=1}^n\,\mu_j=1\,.
\eea
This expression has a particularly simple interpretation in terms of the K\"ahler quotient. Indeed, $\sum\limits_{j=1}^n\,\mu_j\,\log{\mu_j}$ is the symplectic potential of $\CC^n$, and the constraint $\sum\limits_{j=1}^n\,\mu_j=1$ is the restriction of the moment map, corresponding to the diagonal $U(1)$-action. The constraints in~(\ref{GCPN}) define the moment polytope of $\CP^{n-1}$, which is an $(n-1)$-dimensional simplex. 

\vspace{0.3cm}\noindent
Let us now derive, again in the undeformed case, the K\"ahler potential of the $T$-dual metric. As discussed earlier, it is equal to the symplectic potential $H$ of the original metric. Solving the moment map constraint by setting $\mu_k=t_k-t_{k-1}$, where $t_n-t_0=1$, and using~(\ref{GCPN}), we find
\bea\label{KCPNdual}
\mathcal{K}^{\vee}=\sum\limits_{j=1}^n\,(t_j-t_{j-1})\,\log{(t_j-t_{j-1})}\,,\quad\quad\textrm{where}\quad t_j\geq t_{j-1}\,.
\eea
We will be using this formula as a test of our result for the deformed metric in the limit~$\eta\to 0$.

\subsubsection{The `sausage' and its $T$-dual}\label{sausageTdual}
Let us consider in detail the case $n=2$, i.e. the deformation of the sphere $\CP^1\simeq S^2$ that we discussed in Sec.~\ref{Pohlmap}. We start from the metric written in the `stereographic' complex~coordinates
\begin{empheq}[box=\fbox]{align}
\hspace{1em}\vspace{1em}
\label{sausagecompl}
ds^2=
\left({1\over s}-s\right)\,\frac{|dW|^2}{(s+|W|^2)({1\over s}+|W|^2)}\,,\quad\quad\quad 0<s<1\,,
\hspace{1em}
\end{empheq}
which is another form of the sausage metric that may be obtained from~(\ref{2dmetric}), (\ref{gsausage}) by passing to conformal coordinates and setting $s=e^{-\mu_0}$. To bring the metric to the form that was originally found in~\cite{Onofri}, we parametrize $W=e^{{t\over 2}+i \phi}$, in which case ($s^{-1}=e^{\tau}$)
\bea\label{onofrimetr}
ds^2=\sinh{(\tau)}\,\frac{{1\over 4}\,dt^2+d\phi^2}{\cosh(t)+\cosh{(\tau)}}\,,\quad\quad t\in(-\infty, \infty), \;\phi\in[0, 2\pi)\,.
\eea
Interestingly, apart from the $U(1)$ isometry $\phi\to \phi +\mathrm{const.}$, there are two $\mathbb{Z}_2$-isometries: $\phi\to -\phi$ and $t\to -t$. From the point of view of the complex coordinate $W$, they may be combined to the following two isometries: $W\to \bar{W}$ and $W\to {1\over W}$. 

\vspace{0.3cm}\noindent
So far we have encountered several forms of the sausage metric. It is particularly instructive to compare the moment-map form~(\ref{gsausage}) and the conformally flat form~(\ref{onofrimetr}). If one performs a $T$-duality on the $\phi$-angle of the metric~(\ref{gsausage}) and makes a change of variables $\mu_0= \tau+i\,\pi,$ $\mu= t$, one arrives at the metric~(\ref{onofrimetr}). Therefore $T$-duality is effectively performed in this case by a simple analytic continuation of the parameter. Of course this is a formal manipulation, since the range of coordinates also needs to be changed. For example, the range of the $\mu$-coordinate of the $T$-dual metric is in fact infinite, and the metric has singularities as $\mu\to\pm \infty$.

\vspace{0.3cm}\noindent
Let us see, how this analytic continuation works at the level of the K\"ahler potential, which can be read off from the metric~(\ref{onofrimetr}) by comparing to the general toric K\"ahler metric~(\ref{metrtoric1}). It follows that $\mathcal{K}''(t)=\frac{\sinh{(\tau)}}{\cosh(t)+\cosh{(\tau)}}$, so that\footnote{The function $\mathrm{Li}_2(z)$ is the dilogarithm, whose classic definition is $\mathrm{Li}_2(z)=\sum\limits_{n=1}^\infty\,\frac{z^n}{n^2}$, valid for $|z|\leq 1$. It may be analytically continued to $\CC\setminus[1, \infty)$ using the integral expression $\mathrm{Li}_2(z)=-\int\limits_0^z\,\frac{\log{(1-t)}}{t}\,dt$. For details see~\cite{KirillovDilog, Zagier}.}
\begin{empheq}[box=\fbox]{align}
\hspace{1em}\vspace{1em}
\mathcal{K}(W, \bar{W})=\mathrm{Li}_2(-s |W|^2)-\mathrm{Li}_2(-s^{-1} |W|^2),\quad\quad\quad s=e^{-\tau}\,.\hspace{1em}
\end{empheq}
$T$-duality simply flips the sign, $s\to -s$. After this sign flip, as can be seen from~(\ref{sausagecompl}), the range of the coordinates is $|W|^2\in[s, {1\over s}]$. Since $T$-duality is implemented by a Legendre transform at the level of the potential, we come to the following statement: 

\vspace{0.1cm}\noindent
\begin{siderules}
{\large\textbf{Lemma 2.}} The Legendre transform maps $\mathcal{K}_s(t)=\mathrm{Li}_2(-s e^t)-\mathrm{Li}_2(-s^{-1} e^t)$ to $\mathcal{K}_{-s}(t)$.
\end{siderules}

\vspace{0.3cm}\noindent
{\large\textbf{Proof.}} The proof is an application of the five-term identity for the dilogarithm. It is easier to work with Rogers' dilogarithm $L(x)$ defined as $L(x)=\mathrm{Li}_2(x)+{1\over 2}\log{x}\,\log{(1-x)}$ for $0<x<1$. A suitable analytic continuation to the whole real axis is described in~\cite{KirillovDilog}. We start from the so-called Abel identity~\cite{KirillovDilog} $L({1-a\over 1-b})+L({b\over a})=L({1-a\over 1-b}\cdot{b\over a})+L(1-a)+L(b)$. Since $L(a)+L(1-a)=\mathrm{const.}$, we find $L({1-a\over 1-b})+L({b\over a})\simeq L({1-a\over 1-b}\cdot{b\over a})+L(b)-L(a)$. Next we calculate the moment map $\mu-\log{s}={d \mathcal{K}_s(t)\over dt}=\log{({1+s^{-1}e^t\over 1+s e^t})}$ and substitute $a=-s^{-1} e^t, b=-s e^t$ in the identity to find $L(s^{-1}\,e^\mu)+L(s^2)\simeq L(s\, e^\mu)+L(-s e^t)-L(-s^{-1} e^t)$. Returning to the dilogarithm $\mathrm{Li}_2$, we may rewrite this, up to inessential terms that are linear in either $t$ or $\mu$: $t\mu-\mathcal{K}_s(t)\simeq \mathcal{K}_{-s}(\mu)$, which is the definition of the Legendre transform. \dotfill $\blacksquare$

\vspace{0.7cm}\noindent
Let us record for the future, that in the metric~(\ref{sausagecompl}) the limit $s\to 1$ (up to a vanishing factor) corresponds to the round metric on $S^2$, whereas the limit $s\to 0$ corresponds to the flat metric $\frac{|dW|^2}{|W|^2}$ on a cylinder $\mathbb{R}\times S^1$.

\vspace{0.3cm}\noindent
The metric~(\ref{sausagecompl}) has another important regime, where it is still well-defined, namely $s\in \CC$, $|s|=1$. We denote $s=e^{i\tau}$. In this case, when rescaled by a factor of $i$, the metric takes the form $ds^2=\sin{(\tau)}\,\frac{{1\over 4}\,dt^2+d\phi^2}{\cosh(t)+\cos{(\tau)}}$. If we now perform $T$-duality and bring the dual metric to the conformal form, we obtain $(ds^2)^{\vee}=\sin{(\tau)}\,\frac{{1\over 4}\,dt^2+d\phi^2}{\cos(t)-\cos{(\tau)}}$, and the range of the new variable $t$ is $t\in[-\tau, \tau]$. The dual K\"ahler potential is, in this regime,
\bea\label{KunitT}
\mathcal{K}^{\vee}=i\,(\mathrm{Li}_2(e^{i(t+\tau)})-\mathrm{Li}_2(e^{i(t-\tau)})),\quad\quad\quad s=e^{i\tau}\,.
\eea
Strictly speaking, this expression has a non-zero imaginary part. The imaginary part is, however, an elementary function~\cite{Zagier}, and moreover it is linear in $t$, therefore not contributing to the metric. We note that in our generalizations to the case of $\CP^{n-1}$ we will mostly encounter formulas similar to~(\ref{KunitT}), i.e. with the dilogarithm evaluated on the unit circle. Indeed, this is the K\"ahler potential for the metric, $T$-dual to the \emph{unitary} deformation of the $\CP^{n-1}$ model (that is, with $\eta$ real, see~(\ref{qeta})).

\subsection{The $T$-dual K\"ahler geometry for all $n$}\label{TdualKahler}

We pass to the derivation of the $T$-dual model for arbitrary $n$:

\vspace{-0.6cm}\noindent
\begin{siderules}
{\large\textbf{Proposition 2.}} Define the matrix $N_{jk}:=x_j\,(i\,\mathcal{R}_{jk})$ ($j, k =1, \ldots, n$). The metric, $T$-dual to the $\eta$-deformed $\CP^{n-1}$ geometry, has the form:
\bear
\label{defmetricTdual}
&&ds^2=\sum\limits_{j=1}^n\,\frac{d \psi_j^2+\left(dx_j-\eta\,M_j\right)^2}{2 x_j}\,,\\
&&M_j=\sum\limits_{k=1}^n\,\left(N_{jk}-\sum\limits_{i=1}^n\,N_{ik}\,\delta_{jk}\right)\,d\psi_k,\\ \label{defmetricTdual3}
&&\textrm{where}\quad\quad x_j\geq 0,\quad\quad  \sum\limits_{j=1}^n\,x_j=1,\quad\quad \sum\limits_{j=1}^n \,\psi_j=0\,.
\eear
\end{siderules}

\vspace{0.3cm}\noindent
\emph{Comment.} In the limit $\eta\to 0$ the variables $x_j$ become the moment map variables of the original  $\CP^{n-1}$ manifold, albeit viewed as coordinates on the $T$-dual $(\CP^{n-1})^\vee$ manifold. This explains the constraints $x_j\geq 0, \;\sum\limits_j\,x_j=1$ as originating from the analogous constraints for $(\CP^{n-1})$, see~(\ref{GCPN}). Note that these are not the same as the moment map variables of the dual  $(\CP^{n-1})^\vee$, which coincide with the $\tilde{t}$-variables of the original $\CP^{n-1}$. In an attempt to avoid this confusion, we call our variables $x_j$, rather than $\mu_j$.

\vspace{0.7cm}\noindent
{\large\textbf{Proof.}} %
We start from the Lagrangian~(\ref{Lagrdefrho}), parametrizing the variables as $U_j=e^{Y_j}=x_j^{1/2}e^{i\,\phi_j}$. We will also make a change of variables $V_j\to e^{-\bar{Y}_j}\,V_j,\;\;W_j\to e^{-\bar{Y}_j}\,W_j$. The Lagrangian then takes the form
\bear
&&\mathscr{L}=\bar{W}_k\bd Y_k+\bar{V}_k\dd Y_k-W_k\dd \bar{Y}_k-V_k\bd \bar{Y}_k+\\
&&+\sum\limits_{i, \,j}\,\frac{x_j}{x_i}\,(W_i\,\bar{W}_i\,\rho_{ji}+V_i\,\bar{V}_i\,\rho_{ij})+\sum\limits_{i, \,j}\,\rho_{ji}(V_jW_i+\bar{V}_j\bar{W}_i)\,,\\
&&\textrm{and}\quad\quad\sum\limits_{k} W_k=\sum\limits_{k} V_k=0\,.
\eear
Note that the constraints have linearized in the new variables. The constants $\rho_{ij}$ are defined as follows: $\rho(A)=B, B_{ij}=\rho_{ij} A_{ij}$, i.e. $\rho_{ij}=1-i\,\eta\,\epsilon(j-i)$. The Lagrangian is invariant under shifts of the angular variables, $\phi_j\to\phi_j+\alpha_j$. The e.o.m. for $\phi_j$ is simply $\bd(V_k+\bar{W}_k)+\dd(\bar{V}_k+W_k)=0$. According to the standard lore of $T$-duality, we `solve' it by setting
\bea
V_k+\bar{W}_k=i\,\dd\psi_k,\quad\quad \bar{V}_k+W_k=-i\,\bd\psi_k\,,
\eea
where $\psi_k$ are the new (real) variables that are dual to the $\phi$-angles. It is now clear, where the main simplification comes from -- we have gotten rid of half of the variables $(V, W)$ that we needed to `integrate over'. Leaving just the $V$-variables, we obtain the Lagrangian
\bear
&&\mathscr{L}=i\,\sum\limits_k\,\frac{\dd \psi_k \bd x_k+\bd \psi_k \dd x_k}{2\,x_k}+2\,\sum\,\frac{x_j}{x_i}\,|V_i|^2+\sum\,\rho_{ji}\,\frac{x_j}{x_i}\,\dd \psi_i \bd \psi_i+\\ \nonumber
&&+ \sum\limits_k\,V_k\,{1\over x_k}\,\left(-\bd x_k+i\,\left(\sum\limits_j\,x_j\,\rho_{jk}\right)\,\bd \psi_k-i\,x_k\,\sum\limits_j\,\rho_{kj}\,\bd\psi_j+\mathcal{C}\,x_k\right)+\\ \nonumber
&&+ \sum\limits_k\,\bar{V}_k\,{1\over x_k}\,\left( \dd x_k-i\,\left(\sum\limits_j\,x_j\,\rho_{jk}\right)\,\dd \psi_k+i\,x_k\,\sum\limits_j\,\rho_{kj}\,\dd\psi_j-\bar{\mathcal{C}}\,x_k\right) 
\eear
In this derivation we have used the property $\rho_{ij}+\rho_{ji}=2$. Besides, $\mathcal{C}$ is the Lagrange multiplier imposing $\sum\limits_k\,V_k=0$. Upon introducing the new variables $P_i=\sum\limits_j\,\rho_{ji}\,x_j$ and $R:=\sum\limits_j\,x_j$, we eliminate $V_k$:
\bear
&&\!\!\!\!\!\!\!\!\!\!
\scalemath{1}{
\mathscr{L}=i\,\sum\limits_k\,\frac{\dd \psi_k \bd x_k+\bd \psi_k \dd x_k}{2\,x_k}+\sum\limits_i\,{P_i\over x_i}\,\dd \psi_i \bd \psi_i-}\\ \nonumber&&\!\!\!\!\!\!\!\!\!\!
\scalemath{0.92}{
-{1\over 2 R}\,\sum\limits_k\,{1\over x_k}\, \left(- \bd x_k+i\,P_k\,\bd \psi_k-i\,x_k\,\sum\limits_j\,\rho_{kj}\,\bd\psi_j+\mathcal{C}\,x_k\right) \left(\dd x_k-i\,P_k\,\dd \psi_k+i\,x_k\,\sum\limits_j\,\rho_{kj}\,\dd\psi_j-\bar{\mathcal{C}}\,x_k\right)}
\eear
Differentiating w.r.t. $\bar{\mathcal{C}}$, we find $\mathcal{C}=\bd\log{R}$. In fact, using the projective freedom it is convenient to set $R=1$, and therefore $\mathcal{C}=\bar{\mathcal{C}}=0$. In this case 
\bear
&&\mathscr{L}=i\,\sum\limits_k\,\frac{\dd \psi_k \bd x_k+\bd \psi_k \dd x_k}{2\,x_k}+\sum\limits_i\,{P_i\over x_i}\,\dd \psi_i \bd \psi_i-\\ \nonumber&&-{1\over 2 }\,\sum\limits_k\,{1\over x_k}\, \left(- \bd x_k+i\,P_k\,\bd \psi_k-i\,x_k\,\sum\limits_j\,\rho_{kj}\,\bd\psi_j\right) \left(\, \dd x_k-i\,P_k\,\dd \psi_k+i\,x_k\,\sum\limits_j\,\rho_{kj}\,\dd\psi_j\right)
\eear
We can read off the metric  of the target space ($P_k=1+i\,\eta\,\tau_k$):
\bear
&&ds^2=i\,\sum\limits_k\,{d \psi_k d x_k\over  x_k}+\sum\limits_k\,{1+i\,\eta\,\tau_k\over x_k}\,d \psi_k^2+{1\over 2 }\,\sum\,{1\over x_k}\,(dx_k-i\,d\psi_k-\eta\,M_k)^2\\
&&M_k=-\tau_k\,d\psi_k-x_k\,\sum\limits_j\,(i\,\mathcal{R}_{kj})\,d\psi_j\,.
\eear
A direct simplification (getting rid of the fictitious imaginary part) leads to the expression declared in the Proposition.\dotfill $\blacksquare$

\vspace{0.3cm}\noindent
\emph{Comment.} For some purposes it is useful to restore the projective invariance (i.e. $x\to \lambda\,x$) in the metric. Namely, we wish to drop the gauge $\sum\limits_j\,x_j=1$ and work in a gauge-invariant way. In that case the metric may be written as follows:
\bea
ds^2=\sum\limits_{j=1}^n\,\left(\frac{R\,d \psi_j^2}{2 x_j}+\frac{\left(dx_j-x_j\,d\log{R}-\eta\,M_j\right)^2}{2\,R\,x_j}\right),\quad\quad R=\sum\limits_k\,x_k\,.
\eea
This has the local rescaling invariance $x_j \to \lambda x_j$, with $\lambda$ an arbitrary function.

\vspace{0.7cm}\noindent
We come to the description of the K\"ahler structure of the metric:
\begin{siderules}
{\large\textbf{Proposition 3.}} The metric defined in~(\ref{defmetricTdual})-(\ref{defmetricTdual3}) is K\"ahler. In the complex coordinates~$Z_j, j=1, \ldots, n-1$, the K\"ahler potential may be written as ($\eta>0$, $t_0$ and $t_n$ are~parameters)
\bear
\label{KahPotInt}
&&\!\!\!\!\!\!\!\!\!\!\!\!\mathcal{K}=\sum\limits_{j=2}^{n-1}\,i\,\eta(Z_j \bar{Z}_{j-1}-\bar{Z}_j Z_{j-1})+{1\over 2\eta}\,\sum\limits_{j=1}^n\,P\left(2\eta(t_j-t_{j-1})\right)\,,\\ \nonumber 
&&\!\!\!\!\!\!\!\!\!\!\!\!\textrm{where}\quad\quad P(t)=i\left(\mathrm{Li}_2(e^{i t})+{t(2\pi-t)\over 4}\right)\,,\quad\quad   \textrm{and} \quad\quad t_j:=Z_j+\bar{Z}_j  \,.\\ \label{coordrange}
&& \!\!\!\!\!\!\!\!\!\!\!\!\textrm{Coordinate range:} \; -\pi\leq 2\eta\,t_0\leq 2\eta t_1 \leq \cdots \leq 2\eta t_{n-1} \leq 2\eta t_n=-2\eta t_0 \leq \pi
\eear
\end{siderules}

\vspace{0.3cm}\noindent
\emph{Comment.} Before passing to the proof, we perform an elementary check by taking the limit $\eta\to 0$. Note that $P''(t)={1\over 2}\mathrm{ctg}({t\over 2})$, so that in the limit $\eta\to 0$ the asymptotic behavior is $P(t)\to t\log{(t)}$, up to linear terms. The potential then coincides with the potential~(\ref{KCPNdual}) of the undeformed $(\CP^{n-1})^{\vee}$ model.

\vspace{0.7cm}\noindent
{\large\textbf{Proof.}} Let us write the metric in the form $ds^2=\sum\limits_{j=1}^n\,\frac{\theta_j \bar{\theta}_j}{2 x_j}$, where $\theta_j=d \psi_j+i\,(dx_j-\eta\,M_j)$. We will denote $M_j:=\sum\limits_k\,M_{jk}\,d\psi_k$. Due to~(\ref{defmetricTdual3}), the $\theta$-forms are subject to the condition $\sum\limits_{j=1}^n\,\theta_j=0$.

\vspace{0.3cm}\noindent
Now, $\theta_j$ may be taken as a basis of $(0, 1)$-forms of an integrable complex structure. To see this, we need to check that $\Omega^\ast(\mathcal{M})(\theta_1, \ldots, \theta_n)$ is a differential ideal in the ring of forms $\Omega^\ast(\mathcal{M})$. In other words, one needs to show that the $(2,0)$ part of $d\theta_i$ vanishes. We compute $d\theta_j=\eta\,\sum\limits_k\,\mathcal{R}_{jk}(dx_j \wedge d\psi_k+dx_k\wedge d\psi_j)$. Using the fact that $d\psi_j={\theta_j+\bar{\theta}_j\over 2}$ and $dx_j={\theta_j-\bar{\theta}_j\over 2i}+{\eta\over 2}\,\sum\limits_s\,M_{js}\,(\theta_s+\bar{\theta}_s)$, after a simple calculation we find that
\bea
(d\theta_j)^{(2, 0)}=-\frac{i\,\eta^2}{4}\,\sum\limits_{k, s}\,x_s\,\left(\mathcal{R}_{js}\mathcal{R}_{sk}+\mathcal{R}_{jk}\mathcal{R}_{sj}-\mathcal{R}_{jk}\mathcal{R}_{sk}+\delta_{jk}\,\delta_{sk}\right)\,\bar{\theta}_j\wedge \bar{\theta}_k\,.
\eea
We have added the term $\delta_{jk}\,\delta_{sk}$, since it does not affect the result due to the wedge product $\bar{\theta}_j\wedge \bar{\theta}_k$, but it ensures that the expression in brackets is identically $1$. When $j=k=s$, this is due to the delta-symbols. In all other cases, one can decide, which index takes the largest value, which makes it easy to check that the $\mathcal{R}\mathcal{R}$-terms add up to unity. As a result, $(d\theta_j)^{(2, 0)}=-\frac{i\,\eta^2}{4}\,\sum\limits_{k}\,\bar{\theta}_j\wedge \bar{\theta}_k=0$.

\vspace{0.3cm}\noindent
The fundamental Hermitian form, corresponding to the metric, is
\bear\nonumber
\Omega={i\over 2}\sum\limits_{j=1}^n\,\frac{\theta_j \wedge \bar{\theta}_j}{2 x_j}=&&\!\!\!\!\!\!\!\sum\limits_{j=1}^n\,\frac{d\psi_j\wedge(dx_j-\eta\,x_j\,\sum\limits_k (i\,\mathcal{R}_{jk})\,d\psi_k)}{2 x_j}=\\ \label{SympForm}
=&& \!\!\!\!\!\!\!\sum\limits_{j=1}^n\,\frac{d\psi_j\wedge dx_j}{2 x_j}-{\eta\over 2}\,\sum\limits_{j,\,k}\, (i\,\mathcal{R}_{jk})\,d\psi_j\wedge d\psi_k\,.
\eear
Clearly, it is closed: $d\Omega=0$.  

\vspace{0.3cm}\noindent
Next we come to the actual determination of the complex coordinates. In other words, we need to find the integrating multipliers for the set of one-forms $\{\theta_k\}_{k=1 \ldots n}$. We introduce the new coordinates $y_k:=\sum\limits_{i=1}^k\,x_i$ and $\chi_k:=\sum\limits_{i=1}^k\,\psi_i$, as well as the new one-forms $\Theta_k:=\sum\limits_{i=1}^k\,\theta_i$. Note that $y_n=1, \chi_n=0, \Theta_n=0$. Besides, since the $x_k$-variables are non-negative, we find that $y_{k+1}\geq y_k$. It turns out that
\bea\label{tauy}
\Theta_k=(1+i\eta\,(1-2y_k))\,d\chi_k+i\,dy_k\,.
\eea
To prove this, we recall that $(i\mathcal{R})_{jk}=-1$ for $j<k$ and $(i\mathcal{R})_{jk}=1$  for $j>k$. Therefore $\sum\limits_m\,(i\mathcal{R})_{jm} d\psi_m=-\sum\limits_{m\geq j}\,d\psi_m+\sum\limits_{m\leq j}\,d\psi_m=d\chi_{j-1}+d\chi_j$. Besides, we express ${d\psi_k=d\chi_k-d\chi_{k-1}}$, so that
\bear\nonumber
&&\sum\limits_{j=1}^k M_j=\sum\limits_{j=1}^k\,d\chi_j\,(x_j+\sum\limits_{m<j}\,x_m-\sum\limits_{m>j}\,x_m)+\sum\limits_{j=2}^k\,d\chi_{j-1}\,(x_j+\sum\limits_{m>j}\,x_m-\sum\limits_{m<j}\,x_m)=\\ \nonumber &&=-\sum\limits_{j=1}^k\,d\chi_j\,(1-2\,y_j)+\sum\limits_{j=2}^k\,d\chi_{j-1}\,(1-2\,y_{j-1})=-d\chi_k\,(1-2\,y_k)\,.
\eear
Equality~(\ref{tauy}) then follows. It is convenient to make a shift $\chi_k\to \chi_k+{1\over 4\eta} \log{\left(1+\eta^2\,(1-2 \,y_k)^2\right)}$, in which case we get
\bear
&&\Theta_k=(1+i\eta\,(1-2 y_k))\,\left(d\chi_k+i\,\frac{dy_k}{1+\eta^2\,(1-2 y_k)^2}\right)=\\ \nonumber
&&=(1-i\tan{(\eta\,t_k)})\,(d\chi_k+{i\over 2}\,dt_k)=\frac{e^{-i\,\eta\,t_k}}{\cos{(\eta\,t_k)}}\,(d\chi_k+{i\over 2}\,dt_k):=\frac{i\,e^{-i\,\eta\,t_k}}{\cos{(\eta\,t_k)}}\,dZ_k\,,
\eear
where $\tan{(\eta\,t_k)}=\eta\,(2 y_k-1)$ and we have introduced the complex variable $Z_k:={1\over 2}\,t_k-i\,\chi_k$. Due to the constraint $y_{k+1}\geq y_k$ we get $t_{k+1}\geq t_k$. Since $\theta_j=\Theta_j-\Theta_{j-1}$, the metric has the~form
\bea
ds^2=\sum\limits_{j=1}^n\,\eta\,\frac{\|\Theta_j-\Theta_{j-1}\|^2}{\tan{(\eta\,t_j)}-\tan{(\eta\,t_{j-1})}}\,,
\eea
where we assume $\tan{(\eta\,t_{0})}=-\eta$ and $\tan{(\eta\,t_{n})}=\eta$. Therefore in these coordinates the metric is tri-diagonal. Let us first of all analyze the off-diagonal terms. An elementary calculation shows that these terms have the form $-\eta \frac{e^{-i\eta\, (t_j-t_{j-1})}}{\sin{(\eta\,(t_j-t_{j-1}))}} \,dZ_j \,d\bar{Z}_{j-1}$. Let us denote the part of the K\"ahler potential, which describes these terms, by $K(Z_j, \bar{Z}_j, Z_{j-1}, \bar{Z}_{j-1})$. It is not difficult to see that
\bea
K(Z_j, \bar{Z}_j, Z_{j-1}, \bar{Z}_{j-1})=i\,\eta(Z_j \bar{Z}_{j-1}-\bar{Z}_j Z_{j-1})+{1\over 2\eta}\,P(2\eta(t_j-t_{j-1}))\,,
\eea
then $P$ satisfies $P''(t)={1\over 2}\,\mathrm{ctg}{({t\over 2})}$. Next we come to the diagonal terms. A typical diagonal term has the form \bear\nonumber
&&{\eta\over \cos{(\eta\,t_{j})}}\,\left(\frac{\cos{(\eta\,t_{j-1})}}{\sin{(\eta(t_j-t_{j-1}))}}+\frac{\cos{(\eta\,t_{j+1})}}{\sin{(\eta(t_{j+1}-t_{j}))}}\right)\,dZ_j \,d\bar{Z}_{j}=\\ \nonumber&&=\eta\,\left(\mathrm{ctg}{(\eta\,(t_j-t_{j-1}))}+\mathrm{ctg}{(\eta\,(t_{j+1}-t_{j}))}\right)\,dZ_j \,d\bar{Z}_{j}\,.
\eear
These terms come from the differentiation of the `pairwise' potentials $\tilde{K}(Z_j, \bar{Z}_j, Z_{j-1}, \bar{Z}_{j-1})+\tilde{K}(Z_{j+1}, \bar{Z}_{j+1}, Z_{j}, \bar{Z}_{j})$, so there is no extra contribution. As a result, we get expression~(\ref{KahPotInt}) for the full K\"ahler potential  \dotfill $\blacksquare$

\vspace{0.7cm}\noindent
Finally we are in a position to resolve the `paradox' observed in Section~\ref{Tdualtoric}. The question was: how is it possible to get a non-K\"ahler toric manifold after performing $T$-duality on all angles of a toric K\"ahler manifold? Indeed, this would not have been possible if the K\"ahler potential was invariant under the action of the torus (in which case it would have the form~(\ref{Kahtoric})). In the case at hand, however, the potential~(\ref{KahPotInt}) is not invariant under the action of the torus. This is due to the non-invariance of the first term, quadratic in the $Z_j, \bar{Z_j}$, as the imaginary parts of $Z_j$ are the angles. Under a shift of an angular variable, the K\"ahler potential shifts by terms linear in $Z_j, \bar{Z_j}$, which do not contribute to the metric. As a result, the metric is still invariant. A related fact is that the moment map variables are not well-defined, as the $(\CP^{n-1})^\vee$ is not simply-connected. Indeed, a standard argument says that, if a vector field $\xi$ preserves the symplectic form $\Omega$, i.e. $\mathcal{L}_\xi \Omega=0$, it follows that $d(i_\xi \Omega)=0$. If the manifold is simply-connected, or more generally if $[i_\xi \Omega]=0\in H^1(\mathcal{M}, \mathbb{R})$, one can define a moment map $\mu$ by $i_\xi \Omega=d\mu$. In the present case $[i_\xi \Omega]\neq 0\in H^1(\mathcal{M}, \mathbb{R})$, as can be easily seen from the expression~(\ref{SympForm}) for the symplectic form. Indeed $i_{\dd \over \dd\psi_j}\Omega=\frac{ dx_j}{2 x_j}-\eta\,\sum\limits_{k}\, (i\,\mathcal{R}_{jk})\, d\psi_k$, so that the would-be moment map variables $\mu_j$ turn out to be linear in the angles for $\eta\neq 0$ and $n>2$. In this situation the K\"ahler metric inevitably has cross-terms between the angular variables $\psi$ and the radial variables $x$, which produce a non-topological $B$-field upon $T$-duality.

\subsection{Ricci flow}\label{ricciflowsec}
We are now in a position to prove that the $\eta$-deformed metric satisfies the Ricci flow equation, if the parameter $\eta$ is assumed to flow accordingly. To simplify the notation, in the potential~(\ref{KahPotInt}) we make a rescaling $t_j\to {t_j\over 2 \eta}$ and drop the overall factor of $1\over 4\eta$. The dependence on $\eta$ is now only hidden in the dependence of the non-dynamical parameters $t_0, t_n$: $\tan{({t_0\over 2})}=-\eta,\;\tan{({t_n\over 2})}=\eta$. More importantly, we make a shift $t_k\to t_k+2k\tau$, to arrive at the following statement:

\vspace{0.3cm}\noindent
\begin{siderules}
{\large\textbf{Proposition 4.}} The metric described by the K\"ahler potential
\bea\label{metrtau}
\mathcal{K}=\sum\limits_{j=2}^{n-1}\,i\,(Z_j \bar{Z}_{j-1}-\bar{Z}_j Z_{j-1})+2\,\sum\limits_{j=1}^n\,P\left(t_j-t_{j-1}+2\tau\right)\,,
\eea
where $P(t)=i\left(\mathrm{Li}_2(e^{i t})+{t(2\pi-t)\over 4}\right)$, satisfies the Ricci flow equation ${d g_{i\bar{j}}\over d\tau}=4\,R_{i\bar{j}}$\,.
\end{siderules}

\vspace{0.3cm}\noindent
{\large\textbf{Proof.}} For brevity it will be convenient to denote $a_j:=\mathrm{ctg}({t_j-t_{j-1}\over 2}+\tau)$. After the shift, the components of the Hermitian metric are
\bear\nonumber
g_{j\bar{k}}=\underbracket[0.6pt][0.6ex]{(a_{j}+a_{j+1})}_{:=A_j}\delta_{jk}+\underbracket[0.6pt][0.6ex]{(-i-a_{j+1})}_{:=B_j}\,\delta_{j,k-1}+\underbracket[0.6pt][0.6ex]{(i-a_j)}_{:= C_{j-1}}\,\delta_{j,k+1}\,,\quad\quad j, k=1, \ldots, n-1\,.
\eear
The determinant $d_k$ of a tri-diagonal matrix of size $k\times k$ satisfies a well-known recursion relation $d_{k+1}=A_{k+1}\,d_k-B_kC_k\,d_{k-1}$. In our case one computes directly $d_1=\frac{\sin{({t_2-t_0\over 2}+2\tau)}}{\sin{({t_1-t_0\over 2}+\tau)}\sin{({t_2-t_1\over 2}+\tau)}}$ and $d_2=\frac{\sin{({t_3-t_0\over 2}+3\tau)}}{\sin{({t_1-t_0\over 2}+\tau)}\sin{({t_2-t_1\over 2}+\tau)}\sin{({t_3-t_2\over 2}+\tau)}}$. Using elementary trigonometric identities, one can then check that the expression
\bea
d_k=\frac{\sin{\left({t_{k+1}-t_0\over 2}+(k+1)\,\tau\right)}}{\prod\limits_{j=1}^{k+1}\,\sin{\left({t_j-t_{j-1}\over 2}+\tau\right)}}
\eea
satisfies the recursion relation. Therefore the determinant is $\mathrm{det}\,g=d_{n-1}$. Note that $t_0$ and $t_n$ are constant parameters, which leads to the following expression for the Ricci tensor:
\bea
R_{i\bar{j}}=-\,\dd_i\dd_{\bar{j}}\log{(\mathrm{det}\,g)}=\dd_i\dd_{\bar{j}}\left(\sum\limits_{j=1}^n\,\log{\left(\sin{\left({t_j-t_{j-1}\over 2}+\tau\right)}\right)}\right)\,.
\eea
Since $P''(t)={1\over 2}\mathrm{ctg}({t\over 2})$, it follows that $P'(t)=\log{(\sin{t\over 2})}$, so that the Ricci potential is $1\over 4$ times the derivative of the K\"ahler potential w.r.t. $\tau$. The assertion follows immediately. \dotfill $\blacksquare$

\vspace{0.3cm}\noindent
\emph{Comment.} Note that the Ricci flow equation is satisfied without the need for introducing any compensating vector field $\xi^\mu$, or the dilaton $\Phi$. 

\subsubsection{The complex hyperbolic space}
As a preparation for the discussion of what the Ricci flow really does to the metric, let us consider one more limit of the metric~(\ref{metrtau}). We set $\tau=0$ and consider the following limit: $t_0, t_1, \ldots, t_S\to -\pi$, $t_{S+1}, \ldots, t_{n-1}, t_n\to \pi$. Note that since $\tan{t_n \over 2}=\eta$ and $\tan{t_0 \over 2}=-\eta$, this limit implies $\eta\to\infty$. We denote $\pi-t_n:=\epsilon$ and rescale all variables accordingly: $t_j=-\pi+\epsilon \,\tilde{t}_j$ for $j=0, \ldots, S$ and $t_j=\pi+\epsilon \,\tilde{t}_j$ for $j=S+1, \ldots, n$. Taking into account that $t_0=-t_n$, we find $\tilde{t}_0=1$ and $\tilde{t}_n=-1$. The K\"ahler potential takes the form
\bea\nonumber
\scalemath{0.92}{
\mathcal{K}=\epsilon^2\,\sum\limits_{j=1}^n\,{i\over2}\,(\tilde{Z}_j \bar{\tilde{Z}}_{j-1}-\bar{\tilde{Z}}_j \tilde{Z}_{j-1})+\epsilon\,\left(\sum\limits_{\substack{j=1,\\ j\neq S+1}}^n\,(\tilde{t}_j-\tilde{t}_{j-1})\log{(\tilde{t}_j-\tilde{t}_{j-1})}-(\tilde{t}_S-\tilde{t}_{S+1})\log{(\tilde{t}_S-\tilde{t}_{S+1})}\right)}
\eea
(since $\tilde{t}_S\geq 0$ and $\tilde{t}_{S+1}\leq 0$, the difference $\tilde{t}_S-\tilde{t}_{S+1}\geq 0$). Denoting $\mu_j:={1\over 2}(\tilde{t}_j-\tilde{t}_{j-1})$ and dropping the first term as subleading, we get
\bea\label{hypersympot}
\mathcal{K}\sim 2\epsilon\!\!\sum\limits_{\substack{j=1, \\j\neq S+1}}^n\,\mu_j\,\log{|\mu_j|},\quad\quad \sum\limits_{j=1}^n\,\mu_j=-1,\quad\quad \mu_{j\neq S+1}\geq 0, \quad\quad\mu_{S+1}\leq 0 \,.
\eea
This expression coincides with the symplectic potential of the complex hyperbolic space $U(1, n-1)\over U(1)\times U(n-1)$ (compare with an analogous expression~(\ref{GCPN}) for the case of $\CP^n$) and can be easily verified using its K\"ahler potential $\mathcal{K}_0=\log{\left(1-\sum\limits_{j=1}^{n-1}\,|W_j|^2\right)}$. This means that the original $\eta$-deformed manifold (before $T$-duality) coincides, in this limit, with the complex hyperbolic space. In the case of the $\eta$-deformed $\CP^1$ this was observed in~\cite{DMVq}. The factor of $\epsilon$ in front in the formula~(\ref{hypersympot}) indicates that the manifold shrinks in the limit $\epsilon\to 0$ (i.e. $\eta\to\infty$), just like it does in the limit $\eta\to 0$. 

\vspace{0.3cm}\noindent
Note that there are $n$ ways to take the limit that we just described. They correspond to the number of ways to split the times $t_1, \ldots, t_{n-1}$ into two groups, as above, i.e. $S=0, \ldots, n-1$. This is easy to understand at the example of the sphere $\CP^1$, i.e. when $n=2$. In this case one should consider the metric~(\ref{sausagecompl}) in the limit $s\to -1$. Then one arrives at two copies of the Poincar\'e disc, which correspond to the cases $|W|<1$ and $|W|>1$. These are of course connected by the equator $|W|=1$, which serves as the common boundary of the two discs -- it is at infinite distance.

\subsubsection{Endpoints of the Ricci flow}

Earlier we considered the Ricci flow for the K\"ahler metric on a manifold, $T$-dual to the $\eta$-deformed projective space $\CP^{n-1}$. From proposition 4, after shifting the $t$-variables $t_k\to t_k-(2k-n)\tau$, we see that the only effect of the flow is that $t_n=n\tau$ (and $t_0=-n\tau$). Therefore $\eta=\tan{({t_n\over 2})}=\tan{({n\tau\over 2})}$, and the solution interpolates between two endpoints: $t_n=0$ ($\eta\to 0$) and $t_n=\pi$ ($\eta\to\infty$). As the solution approaches $t_n=0$, i.e. when $t_n=\epsilon$ is small and positive, since all the variables satisfy $-\epsilon\leq t_k\leq \epsilon$, one should rescale these variables $t_k\to \epsilon\, t_k$ and take the limit $\epsilon\to 0$. This is the same as taking the $\eta\to 0$ limit in the original potential~(\ref{KahPotInt}). As was described in the Comment after Proposition~3, the resulting geometry is that of the undeformed $(\CP^{n-1})^\vee$. The other limit, $t_n\to\pi$, corresponds to $\eta \to \infty$. The general properties of $T$-duality guarantee that the original $\eta$-deformed geometry also satisfies the Ricci flow equations~\cite{Haagensen}. In order to analyze the implications of the Ricci flow for the original geometry, one should recall that, in the process of performing $T$-duality, a dilaton $\Phi=-{1\over 2}\log{\mathrm{det}({\dd^2 \mathcal{K}\over \dd t^2})}$ is generated~\cite{Buscher}. It contributes to the Ricci flow equation in the original frame, but its effect is fully accounted by the evolution of coordinates according to the gradient flow equation ${d t^k\over d\tau}=-{1\over 2}\,g^{km}\dd_m\Phi$ (cf.~\cite{Oliynyk}). Although we do not demonstrate it here in full generality, we believe that the solutions $t_k(\tau)$ behave, in the limit $t_n=n\tau\to\pi$, as described in the previous section, namely the set $(t_1, \ldots, t_{n-1})$ splits into two groups, the variables from each group approaching the values $-\pi$ and $\pi$, respectively. As a justification of this claim, we consider the case $n=2$, where this effect can already be observed. The $T$-dual metric of the $\eta$-deformed $\CP^1$ is $(ds^2)^{\vee}=\sin{(\tau)}\,\frac{{1\over 4}\,dt^2+d\phi^2}{\cos(t)-\cos{(\tau)}}$ (we encountered it in Sec.~\ref{sausageTdual}). If we perform $T$-duality on~$\phi$, we get the dilaton $\Phi=-{1\over 2}\log{\left(\frac{\sin{\tau}}{\cos(t)-\cos{(\tau)}}\right)}$. The dilaton-induced gradient flow then takes the form ${d t\over d\tau}=-{1\over 2}\,g^{00}\dd_t \Phi={\sin{t}\over \sin{\tau}}$, and as a result $e^\mu:=\frac{\sin({\tau+t\over 2})}{\sin({\tau-t\over 2})}$ is a constant along the flow ($\mu$~is the moment map for the K\"ahler structure). The metric after $T$-duality, in the $\mu$-variable, has the form $ds^2={1\over 4}\left(\,\frac{\sin{(\tau)}}{\cos(t)-\cos{(\tau)}}\right)\,dt^2+\left(\,\frac{\sin{(\tau)}}{\cos(t)-\cos{(\tau)}}\right)^{-1}d\phi^2=\sin{\tau}\,\frac{{1\over 4}d\mu^2+d\phi^2}{\cosh{\mu}+\cos{\tau}}$. This metric satisfies the pure Ricci flow equation ${d{g}_{ij}\over d\tau}={1\over 2}\,R_{ij}$ (just like $(ds^2)^\vee$ in the $t$-variable) and interpolates between the sphere metric at $\tau=0$ and the Lobachevsky metric at $\tau=\pi$. For higher $n$ the $\eta$-deformed model will interpolate, along the Ricci flow, between $\CP^{n-1}$ with its Fubini-Study metric and the complex hyperbolic space $U(1, n-1)\over U(1)\times U(n-1)$ with the Bergman metric, described in the previous section.

\vspace{0.3cm}\noindent
Another possibility is to consider a sausage-like deformation, which means taking $\eta$ imaginary, as explained earlier in Sec.~\ref{sausageTdual}. From the point of view of the K\"ahler potential, it suffices to make a replacement $\eta \to -i\,\eta$. Note however that, after such a replacement, the function $\mathcal{K}$ is no longer real, which reflects the fact that the deformation is not unitary, and the `metric' has an imaginary part (so it should more properly be referred to as a non-degenerate bilinear form on the tangent space). This is the regime studied in~\cite{FateevCP, LitvinovCP}, and within this approach one can analyze the $\sigma$-model as a deformation of a conformal field theory. Just like we did earlier in the unitary case, we rescale the variables $t_k\to {t_k \over 2\eta}$ and, accordingly, $Z_k \to {Z_k \over 2\eta}$, so as to get rid of the parameter $\eta$ in most places. We then get the K\"ahler potential in this regime:
\bear
\label{KahPotIntSausage}
&&\!\!\!\!\!\!\!\!\!\!\!\!\mathcal{K}=\sum\limits_{j=1}^n\,(Z_j \bar{Z}_{j-1}-\bar{Z}_j Z_{j-1})+2\,\sum\limits_{j=1}^n\,\tilde{P}(t_j-t_{j-1})\,,\quad\quad t_j\geq t_{j-1}\,,\\ \nonumber
&&\!\!\!\!\!\!\!\!\!\!\!\!\textrm{where}\quad\quad \tilde{P}(t)=\mathrm{Li}_2(e^{-t})+{t^2\over 4}\,,
\quad\quad   \textrm{and} \quad\quad t_j:=Z_j+\bar{Z}_j  \,.
\eear
One still has the restrictions $t_0 \leq t_1\leq \cdots \leq t_{n-1}\leq t_n$, albeit without an upper bound on~$t_n$. The Ricci flow looks exactly as before, i.e. provided one replaces $t_k\to 2k\tau+t_k$, the K\"ahler potential satisfies the Ricci flow equation in $\tau$. Let us now analyze the endpoints of the Ricci flow, starting with the limit $\tau \to \infty$. In the shifted variables the dilogarithms entering the K\"ahler potential are of the form $\mathrm{Li}_2(e^{-(t_j-t_{j-1}+2\tau)})$, and they all vanish in the limit $\tau\to\infty$, since $\mathrm{Li}_2(x)\sim x$ for $x\to 0$. Besides, the range restriction on the shifted variables is $t_j\geq t_{j-1}-\tau$, so the constraints are lifted in the limit $\tau\to\infty$. Asymptotically, the potential looks as follows (dropping all irrelevant terms):
\bea
\mathcal{K}\sim \sum\limits_{j=1}^n\,(Z_j \bar{Z}_{j-1}-\bar{Z}_j Z_{j-1})+\sum\limits_{j=1}^n\,{(t_j-t_{j-1})^2\over 2}\sim 2\,\sum\limits_{j=1}^n\,(|Z_j|^2-\bar{Z}_j Z_{j-1})\,.
\eea
In this formula $Z_0=Z_n=0$. This is a potential for a flat `metric' on an $(n-1)$-dimensional cylinder $(\CC^\times)^{n-1}$, meaning that the $\sigma$-model is free in this regime, although non-unitary. The limit $\tau\to 0$ of the Ricci flow may be taken exactly as for the unitary case and produces the undeformed space $(\CP^{n-1})^{\vee}$.

\subsection{The $\CP^{n-1}$-model as a $\beta\gamma$-system and its deformation}\label{betagammadef}

\vspace{0.3cm}\noindent
Apart from the deformation described in the previous sections, there is another integrable deformation of the $\CP^{n-1}$-model that can be described by considering the latter as a coupling of two $\beta\gamma$-systems~\cite{CYa}. It is convenient to start with the gauged linear sigma-model (GLSM) formulation of the model, which was constructed in~\cite{BykovNilp}. In the case of the $\CP^{n-1}$-model one starts with three matrices $U\in \mathrm{Hom}(\CC, \CC^n),\, V\in \mathrm{Hom}(\CC^n, \CC),\, \Phi\in \mathrm{End}(\CC^n)$. In other words, $U$ is a column vector and $V$ is a row vector, whereas $\Phi$ is an $n\times n$-matrix. Inspired by~\cite{CYa}, we start from the $\beta\gamma$-system Lagrangian of the following form:
\bea\label{lagr5}
\mathscr{L}=\mathrm{Tr}\left(V \bar{\mathscr{D}} U\right)+\mathrm{Tr}\left(V \bar{\mathscr{D}} U\right)^\dagger+\mathrm{Tr}\left(r_s^{-1}(\Phi_z) \Phi_{\bar{z}}\right)\,,
\eea
where $\bar{\mathscr{D}}$ is the ``elongated'' covariant derivative $\bar{\mathscr{D}} U=\bd U+i \,U \mathcal{A}_{\bar{z}}+i\,\Phi_{\bar{z}} U$
and $r_s$ is the classical $r$-matrix, depending on the deformation parameter $s$, that we already encountered in the previous sections. One can eliminate the auxiliary field $\Phi$ that enters the Lagrangian quadratically. As a result, one arrives at the following:
\bear\label{lagr4}
&&\mathscr{L}=\mathrm{Tr}\left(V \bar{D} U\right)+\mathrm{Tr}\left(V \bar{D} U\right)^\dagger-\mathrm{Tr}\left(r_s(U\otimes V) (U\otimes V)^\dagger\right)\,,\\
&& \textrm{where}\quad\quad \bar{D} U=\bd U+i\, U\mathcal{A}_{\bar{z}}\,.
\eear
Note that the latter is the usual GLSM $\CC^\times$-covariant derivative. We are talking about $\CC^\times$ rather than $U(1)$ as the gauge group, since so far no normalization condition on $U$ is assumed.

\vspace{0.3cm}\noindent
Following~\cite{CYa}, we can construct a family of flat connections by the following procedure. Define 
\bea
\mathscr{A}=r_{\kappa_1}(J_z)\,dz-r_{\kappa_2}(J_{\bar{z}})\,d\bar{z}\,.
\eea
Here $\kappa_1, \kappa_2$ are complex parameters that will be related below, and $J$ is a certain one-form, which in the undeformed case, when $r_s\sim\mathds{1}$, coincides with the Noether current one-form of the model~(\ref{lagr4}):
$
J=J_z\,dz+J_{\bar{z}}\,d\bar{z}=UV\,dz-V^\dagger U^\dagger\,d\bar{z}\,.
$
The connection $\mathscr{A}$ is flat, if two requirements are fulfilled. First, the currents satisfy the equations
\bear
&&\bd J_z=[r_{s^{-1}}(J_{\bar{z}}), J_z]\,,\quad\quad
\dd J_{\bar{z}}=[J_{\bar{z}}, r_s(J_z)]\,.
\eear
These equations follow from the equations of motion of the model~(\ref{lagr4}). Besides, the matrix $r$ has to obey the equation
\bea\label{CYBE1}
r_{\kappa_2}([J_{\bar{z}}, r_s(J_z)])+r_{\kappa_1}([r_{s^{-1}}(J_{\bar{z}}), J_z])+[r_{\kappa_1}(J_z), r_{\kappa_2}(J_{\bar{z}})]=0\,.
\eea
This can be identified with the classical Yang-Baxter equation (CYBE) written out in Appendix~A, if one identifies the parameters as $\kappa_1=e^u,\; \kappa_2=e^{u+v},\; e^v=s^{-1}$. In particular, this implies $\kappa_1=\kappa_2\,s\,.$ In the trigonometric case the $r$-matrix satisfying this equation is given by~(\ref{trigrmatrix}). 

\vspace{0.3cm}\noindent
The deformation~(\ref{lagr4}) is generally different from the $\eta$-deformation that we considered earlier. The case where the two deformations coincide is $n=2$, i.e. when we are again in the situation of the deformed sphere $\CP^1$. The reason why the Lagrangian~(\ref{Lagrdefrho}) in this case leads to the same theory is that its last line vanishes for $n=2$. Indeed, consider the term $\mathrm{Tr}(U\otimes \bar{V}\,\rho(U\otimes \bar{W}))=\sum\limits_{i, j}\,U_i \bar{V}_j U_j \bar{W}_i (1-\eta\,\mathcal{R}_{ji})=-{\eta \over 2}\,\sum\limits_{i, j}\,U_i U_j \,\mathcal{R}_{ji}\,(\bar{V}_j  \bar{W}_i-\bar{V}_i  \bar{W}_j )$, where we took into account that the vectors $U$ and $V$ are orthogonal, $\bar{V}\circ U=0$. The case $n=2$ is special for the following reason: since both $V$ and $W$ are orthogonal to $U$, they have to be collinear, thereby leading to the vanishing of the combination in brackets in the above formula. Integrating out $V$ and $W$ from the remaining terms will produce, after some simple rewritings, a model equivalent to the one of~(\ref{lagr4}). In fact, the authors of~\cite{CYa} used, in their Section 6.3, exactly the formalism of~(\ref{lagr4}) to derive the sausage model.

\vspace{0.3cm}\noindent
On the other hand, the model~(\ref{lagr4}) for $n>2$ is subject to the so-called Pontryagin anomaly~\cite{NekrasovBeta, Witten02}, as was also mentioned in~\cite{CYa}. The anomaly is governed by the fact that $p_1(\CP^{n-1})\neq 0$ for $n>2$. Therefore it is unclear, whether the deformation of this type, or some modification thereof, may be defined at the quantum level. Questions about the $\beta$-function, Ricci flow, etc. need to be postponed until these issues are fully settled.

\section{Flag manifold models and Toda field theories}\label{Todasec}

In the previous sections we discussed in detail the $\eta$-deformation of $\CP^{n-1}$ target spaces. It would be desirable to eventually extend those results to the case of generic flag manifold target spaces. As a first step in this direction, we turn to the analysis of the Pohlmeyer map for complete flag manifolds
\bea\label{completeflag}
\mathcal{M}=\frac{U(n)}{U(1)^n}\,,
\eea
which, in a sense, is the case polar to the one of symmetric target spaces. A point in this manifold is an $n$-tuple of ordered orthonormal vectors in $\CC^n$: $(u_1, \ldots, u_n):=g$, which we have grouped into a unitary matrix $g$. It is customary to introduce a Maurer-Cartan current $J=g^{-1}dg$ that satisfies the zero-curvature equation $dJ+J\wedge J=0$. Let us also introduce the notation $G=U(n), \;H=U(1)^n$, the corresponding Lie algebras being $\mathfrak{g}$ and $\mathfrak{h}$. Now,
\bear
&&\mathfrak{g}=\mathfrak{h}\oplus \mathfrak{m}\quad\quad\quad\;\;\; (\mathfrak{m}= \textrm{off-diagonal matrices})\\
&&\mathfrak{m}=\mathfrak{m}_+\oplus \mathfrak{m}_-\quad\quad (\mathfrak{m}_\pm= \textrm{upper/lower-triangular matrices})
\eear
We can decompose the current accordingly: $J=J_0+J_++J_-$. We will take the $\sigma$-models of~\cite{BykovNon, BykovSols, BykovZeroCurv} as the relevant generalizations of symmetric space $\sigma$-models. As shown in~\cite{BykovZeroCurv}, the e.o.m. of these  models may be concisely written as (here $\mathscr{D}$ is the $H$-covariant derivative)
\bear\label{eom1}
\bar{\mathscr{D}}(J_+)_z-[(J_+)_z, (J_+)_{\bar{z}}]=0\,.\\ \label{eom2}
\mathscr{D}(J_-)_{\bar{z}}+[(J_-)_z, (J_-)_{\bar{z}}]=0\,.
\eear
The projection of the zero-curvature condition $dJ+J\wedge J=0$ onto $\mathfrak{m}_-$ has the form
\bea\label{eom11}
\mathscr{D}(J_-)_{\bar{z}}-\bar{\mathscr{D}}(J_-)_z+[(J_-)_z, (J_-)_{\bar{z}}]+[(J_+)_z, (J_-)_{\bar{z}}]_{\mathfrak{m}_-}+[(J_-)_z, (J_+)_{\bar{z}}]_{\mathfrak{m}_-}=0\,.
\eea
Taking into account~(\ref{eom1}), we obtain
\bea\label{eom3}
\bar{\mathscr{D}}(J_-)_z-[(J_+)_z, (J_-)_{\bar{z}}]_{\mathfrak{m}_-}-[(J_-)_z, (J_+)_{\bar{z}}]_{\mathfrak{m}_-}=0\,.
\eea
We can now derive an important conservation law. First of all, (\ref{eom1}) implies that $\bar{\mathscr{D}}(J_+)_z\in [\mathfrak{m}_+, \mathfrak{m}_+]$, therefore $\bar{\mathscr{D}}(J_z)_{i,i+1}=0, \;\;i=1, \ldots n-1$. Besides, (\ref{eom3}) implies that $\bar{\mathscr{D}}(J_-)_z\in \mathfrak{m}_-\cap [\mathfrak{m}_+, \mathfrak{m}_-]$, so that $\bar{\mathscr{D}}(J_z)_{n, 1}=0$. This leads to the conservation law
$
\bd\left(\prod\limits_{i=1}^n\,(J_z)_{i, i+1} \right)=0\,,
$
where we have assumed a cyclic labeling of the currents, $J_{n, n+1}\equiv J_{n, 1}$. Using conformal invariance and assuming that $(J_{i, i+1})_z\neq 0$ for all $i$, we may set
\bea\label{cyclicprod}
\prod\limits_{i=1}^n\,(J_z)_{i, i+1}=1\,.
\eea
This is a generalization to arbitrary $n$ of Pohlmeyer's constraint~(\ref{pohlunitcond}). The constraint can be `solved' by setting
\bea\label{exppar}
(J_{i, i+1})_z=e^{X_{i}-X_{i+1}},\quad\quad X_i\in \CC\quad\textrm{mod}\quad 2\pi i\,\mathbb{Z}\,.
\eea
Note that the definition of the flag manifold implies a gauge symmetry $u_k\to e^{i\alpha_k} \,u_k$. In terms of the variables $X_i$, this is a shift symmetry $X_k\to X_k-i\,\alpha_k$. In particular, a suitable gauge is taken by assuming that these variables are real: $X_k\in \mathbb{R}\,.$ In what follows this choice of gauge will be implied.

\vspace{0.3cm}\noindent
The equations $\bar{\mathscr{D}}(J_z)_{i,i+1}=0$ may now be solved by identifying $\bar{u}_i \circ \bd u_i=-\bd X_i$. In other words, in the new variables the connection $A_k=i \,J_{kk}=i \bar{u}_k \circ d u_k$ entering the covariant derivative $\mathscr{D}$ may be written as
\bea\label{Akconn}
A_k=i \,J_{kk}=i (\dd X_k \,dz-\bd X_k \,d\bar{z})\,.
\eea

\subsection{Toda parametrization}

In this section we consider the simplest (non-symmetric) flag manifold $U(3)\over U(1)^3$. We will assume that ${\prod\limits_{i=1}^3\,(J_z)_{i, i+1} \neq 0}$, so that~(\ref{cyclicprod}) may be achieved by a conformal transformation\footnote{As it was shown in~\cite{BykovSols}, the case $\prod\limits_{i=1}^3\,(J_z)_{i, i+1}\equiv 0$ effectively reduces the e.o.m. to those of the $\CP^2$ $\sigma$-model.}. 

\vspace{0.3cm}\noindent
The expression for $J_z$ in the $X$-variables has the form:
\bea
J_z=
\begin{pmatrix}
\dd X_1 & e^{X_1-X_2}& (J_{13})_z\\
(J_{21})_z & \dd X_2 & e^{X_2-X_3}\\
e^{X_3-X_1}& (J_{32})_z & \dd X_3
\end{pmatrix},\quad\quad\quad J_{\bar{z}}=-J_z^\dagger\,.
\eea
$(J_+)_z$ and $(J_-)_z$ are respectively the upper- and lower-triangular parts of the above matrix, and $(J_+)_{\bar{z}}=-((J_-)_z)^\dagger$, $(J_-)_{\bar{z}}=-((J_+)_z)^\dagger$. To write down the equations, we introduce the `twisted currents' $(U_{21})_z=e^{X_1-X_2} (J_{21})_z, (U_{13})_z=e^{X_3-X_1} (J_{13})_z, (U_{32})_z=e^{X_2-X_3} (J_{32})_z$. The first equation~(\ref{eom1}) then gives:
\bea
\bd(U_{13})_z+e^{2(X_3-X_2)}\,\widebar{(U_{32})_z}-e^{2(X_2-X_1)}\,\widebar{(U_{21})_z}=0\,.
\eea
The remaining equations, which follow from~(\ref{eom3}), are
\bear
&&\bd(U_{21})_z+e^{2(X_1-X_3)}\,\widebar{(U_{13})_z}-e^{2(X_3-X_2)}\,\widebar{(U_{32})_z}=0\,,\\
&&\bd(U_{32})_z-e^{2(X_1-X_3)}\,\widebar{(U_{13})_z}+e^{2(X_2-X_1)}\,\widebar{(U_{21})_z}=0
\eear
The equations imply the conservation law $\bd((U_{13})_z+(U_{21})_z+(U_{32})_z)=0$, where $T_{zz}:=(U_{13})_z+(U_{21})_z+(U_{32})_z$ is the holomorphic component of the energy-momentum tensor.

\vspace{0.3cm}\noindent
One still has the equations arising from the $\mathfrak{h}$-projection of the zero-curvature equation ${dJ+J\wedge J=0}$:
\bea \nonumber
\!\!\!\!\!\!\!\!\!\!\!\!\!\!\!\!\!\!\!\!2 \,\dd \bd X_1+e^{2(X_1-X_2)}-e^{2(X_3-X_1)}+e^{2(X_1-X_3)}\|(U_{13})_z\|^2-e^{2(X_2-X_1)}\|(U_{21})_z\|^2=0\eea\bear 
&&\!\!\!\!\!\!\!\!\!\!\!\!\!\!\!\!\!\!\!\!2 \,\dd \bd X_2+e^{2(X_2-X_3)}-e^{2(X_1-X_2)}+e^{2(X_2-X_1)}\|(U_{21})_z\|^2-e^{2(X_3-X_2)}\|(U_{32})_z\|^2=0\\ \nonumber
&&\!\!\!\!\!\!\!\!\!\!\!\!\!\!\!\!\!\!\!\!2 \,\dd \bd X_3+e^{2(X_3-X_1)}-e^{2(X_2-X_3)}+e^{2(X_3-X_2)}\|(U_{32})_z\|^2-e^{2(X_1-X_3)}\|(U_{13})_z\|^2=0\,.
\eear
These are the generalized Toda equations -- a system describing the interaction of the Toda fields with the additional $U$-fields. In particular, if one sets $(U_{13})_z=(U_{21})_z=(U_{32})_z=0$, one obtains the pure Toda equations. Note that since the equations for $U_{ij}$ are of first order, a curious prospect would be in devising a fermionic interpretation for these variables.

\vspace{0.3cm}\noindent
In this section we considered in detail the case $n=3$, but the results are generalizable to arbitrary $n$, that is, to flag manifolds of the type~(\ref{completeflag}). It would also be interesting to construct an analogous parametrization for flag manifolds of arbitrary type, i.e. allowing for non-abelian denominators.

\subsection{Toda reduction in the general case}\label{genTodasec}

In this section we will derive a Toda field theory from a flag manifold $\sigma$-model for general~$n$. In doing so, we will restrict ourselves to the pure Toda field theory, i.e. when all the additional $U$-fields are set to zero. For general $n$, this condition may be formulated, using the $\mathbb{Z}_n$-grading on the Lie algebra $\mathfrak{g}\simeq\mathfrak{su}_n$. The latter means a decomposition $\mathfrak{g}=\oplus_{i=0}^{n-1}\,\mathfrak{g}_i$, such that $[\mathfrak{g}_i, \mathfrak{g}_j]\subset \mathfrak{g}_{i+j\;\;\mathrm{mod}\;\; n}$. The subspaces $\mathfrak{g}_i$ are in fact uniquely determined by $\mathfrak{g}_1$, and we define $\mathfrak{g}_1=\mathrm{Span}(E_{i, i+1})_{i=1, \ldots n}$, where $E_{i, i+1}$ are the elementary `unit'-matrices, and $E_{n, n+1}\equiv E_{n, 1}$. It follows that $\mathfrak{g}_{-1}=\mathrm{Span}(E_{i+1, i})_{i=1, \ldots n}$ (again $E_{n+1, n}\equiv E_{1, n}$). The reduction to zero $U$-fields is then formulated as follows:
\bea\label{primred}
J_z\in \mathfrak{g}_0\oplus\mathfrak{g}_1,\quad\quad J_{\bar{z}}\in \mathfrak{g}_0\oplus\mathfrak{g}_{-1}\;.
\eea
Therefore $[(J_+)_z, (J_-)_{\bar{z}}]\in \mathfrak{h},\, [(J_-)_z, (J_+)_{\bar{z}}]\in \mathfrak{h}$, and the equations (\ref{eom1}), (\ref{eom3}) reduce to $\mathscr{D}_{\bar{z}}(J_{i, i+1})_z=0$, which are automatically solved by~(\ref{exppar}),~(\ref{Akconn}). The remaining non-trivial equations are the diagonal components of~$dJ+J\wedge J=0$, and these are the Toda chain equations
\bear
&&2\,\dd\bd X_k-[J_z, J_{\bar{z}}]_{kk}=\\ \nonumber
&&=2\,\dd\bd X_k+e^{2(X_k-X_{k+1})}-e^{2(X_{k-1}-X_k)}=0\,.
\eear
In the literature on harmonic maps, cf.~\cite{Bolton, Guest}, it has been observed that these Toda equations arise in the context of the so-called $n$-primitive maps to flag manifolds. These are maps $g: \Sigma\to {U(n)\over U(1)^n}$, whose associated current $J=g^{-1}dg$ satisfies the constraints~(\ref{primred}). The virtue of our approach is that we have a full theory, of which the Toda chain is a subsector that can be described by simple constraints.

\vspace{0.3cm}\noindent
To illustrate the last point, let us derive the Lax pair of Flaschka-Manakov type from the Lax pair of the flag manifold $\sigma$-model. We start from the flag manifold generalization of Pohlmeyer's family of flat connections~(\ref{pohlflat1}) discussed earlier in the context of the \mbox{$\CP^1$-model:}
\bear
&&A_u={1-u\over 2} K_z dz+{1-u^{-1}\over 2} K_{\bar{z}} d\bar{z}\,,\quad\textrm{where}\\
&& K_z=g\begin{pmatrix}
0  & 2 \,(J_{12})_z  & 0 & \cdots  & 0 \\
0  & 0  & 2\,(J_{23})_z  & 0  &  \vdots \\
\vdots & \ddots & \ddots & \ddots     & \vdots \\
0 & \cdots &  \cdots & 0 &    2 \,(J_{n-1, n})_z  \\
0 & \cdots &  \cdots & \cdots &    0 
\end{pmatrix}g^\dagger,\quad\quad K_{\bar{z}}=-K_z^\dagger
\eear
Performing a gauge transformation by the group element $g$ and introducing the $X$-variables according to~(\ref{exppar}), (\ref{Akconn}), we obtain
\bear
&&\!\!\!\!\!\!\!\!\!\!\!\!\!\!\!\mathscr{A}_u=\\ \nonumber&&\!\!\!\!\!\!\!\!\!\!\!\!\!\!\!\scalemath{0.94}{\begin{pmatrix}
\bd X_1\,d\bar{z}-\dd X_1 \,dz  & -u \,e^{X_1-X_2}\,dz  & 0 & \cdots  & e^{X_n-X_1}\,d\bar{z} \\
u^{-1} \,e^{X_1-X_2}\,d\bar{z}  & \bd X_2\,d\bar{z}-\dd X_2 \,dz   & -u\,e^{X_2-X_3}\,dz  & 0  &  \vdots \\
0 & \ddots & \ddots & \ddots     & \vdots \\
\vdots & \ddots & \ddots & \ddots     & \vdots \\
0 & \cdots &  \ddots & \bd X_{n-1}\,d\bar{z}-\dd X_{n-1} \,dz  &    -u \,e^{X_{n-1}-X_n}\,dz  \\
-e^{X_n-X_1}\,dz & 0 &  \cdots & u^{-1} \,e^{X_{n-1}-X_n}\,d\bar{z} &    \bd X_n\,d\bar{z}-\dd X_n \,dz 
\end{pmatrix}}
\eear
$\mathscr{A}_u$ furnishes a one-parameter family of flat connections, describing Toda field theory. One can also perform a mechanical reduction by postulating $\bd =\dd ={d\over dt}$, in which case one is led to the mechanical Toda chain~\cite{Toda}. If one additionally sets $u=1$, the zero-curvature equation (which in this case coincides with the flatness condition for $J$) takes the form discovered by Manakov~\cite{Man} (in a different form by Flaschka~\cite{Fl})
\vspace{-0.2cm}
\bea
\dot{L}=[L, M],\quad\quad \textrm{where}\quad\quad L=(\mathscr{A}_{1})_{\bar{z}}-(\mathscr{A}_{1})_{z},\quad M={1\over 2}\left((\mathscr{A}_{1})_{\bar{z}}+(\mathscr{A}_{1})_{z}\right)\,.
\eea
For a review of mechanical Toda chains and their Lax representations one can consult~\cite{Perelom}.

\section{Conclusion and outlook}

Let us summarize the main findings of the present paper, outlining some open problems and directions for future research. First, in section~\ref{Pohlmap} we considered a $\sigma$-model with a $U(1)$-invariant target space, constructed its `Pohlmeyer-reduced' cousin and showed that the requirement of its integrability naturally leads, in a unique way, to the so-called sausage metric, first considered in~\cite{Onofri}. It would be interesting to generalize this argument to higher-dimensional target spaces, for example to the case of the deformed $\CP^{n-1}$ model.

\vspace{0.3cm}\noindent
In section~\ref{etadefsec} we studied the $\eta$-deformed $\CP^{n-1}$ model, starting from `first principles'. This manifold is toric, and it turned out useful to pass to the $T$-dual frame, i.e. to dualize all the angles simultaneously. Whereas the original geometry is generalized K\"ahler~\cite{DemulderCP}, we proved that the dual geometry is K\"ahler and derived its K\"ahler potential in explicit form. It would be interesting to derive the generalized K\"ahler potential for any $n$ as well (the cases $n=2$ and $n=3$ were studied in~\cite{DemulderCP}). At least when the K\"ahler potential is invariant under the action of the torus, the dual potential is given by a Legendre transform~\cite{RocekVerlinde}. As explained at the end of section~\ref{TdualKahler}, in our case the K\"ahler potential is not invariant, despite the fact that the metric is. On the other hand, the expression in~\cite{DemulderCP} for the generalized K\"ahler potential at $n=3$ is strikingly similar to our expression~(\ref{KahPotInt}), which is reminiscent of the phenomenon described in our lemma~2.

\vspace{0.3cm}\noindent
We then showed that the $\eta$-deformed metric satisfies the K\"ahler Ricci flow equation in the $T$-dual frame. The original geometry then interpolates, along the Ricci flow, between the $\CP^{n-1}$ manifold with its Fubini-Study metric and the complex hyperbolic space $U(1, n-1)\over U(1)\times U(n-1)$ with the Bergman metric. One can also achieve an `asymptotically free' regime, if one chooses the deformation parameter to be imaginary.

\vspace{0.3cm}\noindent
Finally, in section~\ref{Todasec} we turned to the classically integrable $\sigma$-models with flag manifold target spaces, introduced by one of the authors~\cite{BykovNon, BykovSols, BykovZeroCurv}. We considered the class of complete flag manifolds and demonstrated that, if a certain non-degeneracy condition~(\ref{cyclicprod}) holds, the model may be rewritten as a Toda field theory interacting with additional fields (which we call $U_{ij}$). The fact that the Toda chain arises in the context of harmonic maps into flag manifolds (the so-called `primitive maps') has been known for a while~\cite{Bolton, Guest}, however in section~\ref{Todasec} we derive this as a reduction ($U_{ij}=0$) of a complete non-linear $\sigma$-model. The mapping of the flag manifold $\sigma$-model to the Toda chain is akin to Pohlmeyer's reduction~\cite{Pohlmeyer, Eichenherr, Grigoriev, Miramontes}, and it would be interesting to extend this map to the case of generic flag manifolds -- a vast class ranging between complete flag manifolds and Grassmannians.

\vspace{0.3cm}\noindent
\textbf{Acknowledgments.} D.B. would like to thank A.~A.~Slavnov for support. We would like to thank M.~Alfimov, S.~Demulder, S.~Frolov and A.~Litvinov for valuable discussions.
The work of D.L. is supported  by the Origins Excellence Cluster.

\appendix \subsubsection{The classical Yang-Baxter equation}\label{CYBEapp}

In its conventional form, the classical Yang-Baxter equation involves the classical $r$-matrix $r(u)\in \mathfrak{g}\otimes\mathfrak{g}$. The equation itself takes values in $\mathfrak{g}\otimes\mathfrak{g}\otimes\mathfrak{g}$ and has the following form:
\bea
[r_{12}(u), r_{13}(u+v)]+[r_{12}(u), r_{23}(v)]+[r_{13}(u+v), r_{23}(v)]=0\,.
\eea
Here we write the equation in additive form, i.e. when the spectral parameters $u$ and $v$ take values in $\CC$. The notation $r_{12}(u)$ means $r_{12}(u)=r(u)\otimes \mathds{1}$, and analogously for other pairs of indices. Solutions to the above equation have been extensively investigated in the classical paper of Belavin and Drinfeld~\cite{BD}.

\vspace{0.3cm}\noindent
For the purposes of the present paper it is more convenient to think of the $r$-matrix as a map $r: \mathfrak{g} \to \mathfrak{g}$, or equivalently $r\in \mathrm{End}(\mathfrak{g})\simeq \mathfrak{g}\otimes \mathfrak{g}^\ast$. To rewrite the equation accordingly, first of all we expand the $r$-matrix in the basis of generators $\tau_a \in \mathfrak{g}$:
\bea
r(u)=\sum\limits_{i,\,j=1}^{\mathrm{dim}\,\mathfrak{g}}\,\lambda_{ij}(u)\,\tau_i\otimes \tau_j\,.
\eea
The three terms in the equation can then be written as (we assume that the generators form an orthonormal basis w.r.t. the Killing metric, and $[\tau_a, \tau_b]=f_{abc}\,\tau_c$)
\bear
&& [r_{12}(u), r_{13}(u+v)]=\sum\,\lambda_{ij}(u) \lambda_{mn}(u+v) [\tau_i, \tau_m]\otimes \tau_j\otimes \tau_n\\
&&[r_{12}(u), r_{23}(v)]=\sum\,\lambda_{ij}(u) \lambda_{mn}(v)\,f_{jms}\,\tau_i\otimes \tau_s \otimes \tau_n\\
&&[r_{13}(u+v), r_{23}(v)]=\sum\,\lambda_{ij}(u+v) \lambda_{mn}(v)\,f_{jns}\,\tau_i\otimes \tau_m\otimes \tau_s
\eear
We will now drop the index $12, 13, 23$-notation and write $r_u(a)\in \mathfrak{g}$ for the $r$-matrix acting on a Lie algebra element $a\in \mathfrak{g}$. Inserting the Lie algebra elements $a, b$ in the second and third slots of the equation, we obtain
\bear
&&[r_{12}(u), r_{13}(u+v)](a, b)=[r_u(a), r_{u+v}(b)]\\
&&[r_{12}(u), r_{23}(v)](a, b)=r_u([r_v(b), a])\\
&&[r_{13}(u+v), r_{23}(v)](a, b)=r_{u+v}([b, r_{-v}(a)])
\eear
To derive the last equality, we have used the assumed `unitarity' property of the $r$-matrix: $\lambda_{ji}(u)=-\lambda_{ij}(-u)$. As a result, the CYBE takes the form
\bea
[r_u(a), r_{u+v}(b)]+r_u([r_v(b), a])+r_{u+v}([b, r_{-v}(a)])=0\,.
\eea

\makeatletter
\renewcommand\@biblabel[1]{#1.}
\makeatother

{ \setlength{\bibsep}{0.0pt}

  }

\end{document}